\documentclass[preprint,amsmath,amssymb,floatfix,longbibliography,superscriptaddress]{elsarticle}

\usepackage{graphicx}
\usepackage{amsmath,amssymb,bbold,bm,color}
\usepackage{float}
\usepackage[caption=false, justification=raggedright]{subfig}
\usepackage{epstopdf}
\usepackage{hyperref}
\usepackage[dvipsnames]{xcolor}
\usepackage[normalem]{ulem} 
\usepackage{comment}

\usepackage{array}
\newcolumntype{P}[1]{>{\centering\arraybackslash}p{#1}}

\newcommand{\bra}[1]{\langle #1 |}
\newcommand{\ket}[1]{| #1 \rangle}

\newcommand{\Vr}{{\bm r}}
\newcommand{\VbR}{\bm{R}}
\newcommand{\bA}{\bm{A}}
\newcommand{\bG}{\bm{G}}
\newcommand{\bB}{\bm{B}}
\newcommand{\bv}{\bm{v}}
\newcommand{\bu}{\bm{u}}
\newcommand{\FranzTesanovic}{Franz-Te\ifmmode \check{s}\else \v{s}\fi{}anovi\ifmmode \acute{c}\else \'{c}\fi{} }

\graphicspath{{Figures/}}

\hypersetup{colorlinks=true,linkcolor=blue,citecolor=blue,urlcolor=blue}

\begin{document}

\title{Majorana bound states in vortex lattices on iron-based superconductors}

\author[UBC]{Vedangi Pathak}
\author[UBC]{Stephan Plugge\corref{corrauth}}
\author[UBC]{Marcel Franz}

\cortext[corrauth]{Corresponding author: \texttt{plugge@phas.ubc.ca}}
\address[UBC]{Department of Physics and Astronomy \& Stewart Blusson Quantum Matter Institute, University of British Columbia, Vancouver BC, Canada V6T 1Z4}

\date{\today}

\begin{abstract}
Majorana quasi-particles may arise as zero-energy bound states in vortices on the surface of a topological insulator that is proximitized by a conventional superconductor.
Such a system finds its natural realization in the iron-based superconductor FeTe$_{0.55}$Se$_{0.45}$ that combines bulk $s$-wave pairing with spin helical Dirac surface states, and which thus comprises the ingredients for Majorana modes in absence of an additional proximitizing superconductor.
In this work, we investigate the emergence of Majorana vortex modes and lattices in such materials depending on parameters like the magnetic field strength and vortex lattice disorder.
A simple 2D square lattice model here allows us to capture the basic physics of the underlying materials system. To address the problem of disordered vortex lattice, which occurs in real systems, we adopt the technique of  the singular gauge transformation which we modify such that it can be used in a system with periodic boundary conditions.
This approach allows us to go to larger vortex lattices than otherwise accessible, and is successful in replicating several experimental observations of Majorana vortex bound states in the FeTe$_{0.55}$Se$_{0.45}$ platform. Finally it can be related to a simple disordered Majorana lattice model that should be useful for further investigations on the role of interactions, and towards topological quantum computation.

\end{abstract}


\date{\today}
\maketitle

\section{Introduction}

Majorana fermions are particles that are their own anti-particle. In the context of condensed matter physics, where electrons and holes serve as the canonically defined particles and anti-particles, Majorana fermions arise as non-trivial emergent excitations~\cite{Alicea2012review,Leijnse2012review,Beenakker2013review,Franz2015review}.
Majorana quasi-particles provide insight into exotic fundamental physics along with exciting technological applications in topological quantum computation~\cite{Lutchyn2018review,Prada2020review,Beenakker2019review}. Early on, it was proposed that spinless $p$-wave superconductors can host Majorana quasi-particles~\cite{Kitaev2001,ReadGreen2000,Ivanov2001}. A one-dimensional $p$-wave superconductor exhibits a bulk topological phase and hosts Majorana zero-energy modes at its boundaries~\cite{Kitaev2001}. In a two-dimensional spinless $p_{x}+\text{i}p_{y}$ superconductor, Majorana bound states occur at the centers of half-flux vortices~\cite{ReadGreen2000,Ivanov2001}.
However, low-dimensional $p$-wave superconductivity is extremely rare in nature which makes the detection and control of Majorana quasi-particles in such systems correspondingly more difficult. The seminal Fu-Kane proposal~\cite{Fu2008} showed that the proximity effect between an ordinary $s$-wave superconductor (SC) and the surface states of a strong topological insulator (TI) leads to formation of a state that also hosts the zero-energy Majorana bound states at vortices.

The iron-based superconductor FeTe$_{0.55}$Se$_{0.45}$ has piqued a growing recent interest, owing to the presence of spin helical Dirac surface states with a large superconducting gap induced by its SC bulk~\cite{Zhang2018,Wang2018,Machida2019,Kong2019,Kong2020}. These features make it a natural realisation of the Fu-Kane model~\cite{Fu2008}, without the need to create a TI-SC heterostructure, and pave the way to observe zero-energy Majorana bound states (MBS) in vortices threading its surface.
A distinguishing characteristic of this system is its low Fermi energy, which ensures that the MBS - if present - remain relatively isolated from the higher-energy in-gap Caroli-de Gennes-Matricon (CdGM) bound states~\cite{Caroli1964}.
Several recent scanning tunneling microscopy (STM) experiments indeed have revealed zero-bias peaks and other in-gap modes in FeTe$_{0.55}$Se$_{0.45}$~\cite{Wang2018,Chen2018,Machida2019,Kong2019,Kong2020}.
Interestingly, the experiments found intriguing dependencies of the vortex tunneling spectra on magnetic field, temperature, and disorder, in contrast with the general expectation for a stable zero-bias peak due to the vortex' MBS. Understanding these results better will provide insight into the emergent properties of multiple MBSs in vortex lattices, and is crucial towards the control and manipulation of Majorana quasi-particles for braiding and topological quantum computing.

The dependence of zero-bias peaks on the applied magnetic field has been the topic of contention among several experimental studies~\cite{Wang2018,Chen2018}. Ref.~\cite{Chen2018} reported the absence of zero-modes at high magnetic fields, while another group~\cite{Wang2018} observed zero-bias peaks but only for about 20\% of the vortices and at a lower magnetic field.
A recent comprehensive, high-resolution STM study~\cite{Machida2019} exploring the dependence of Majorana zero-modes on magnetic field, ascertained that zero-bias peaks are not uniformly found in all vortices. However, the observed variation in vortex spectra appeared to be uncorrelated with disorder in the chemical composition or electronic structure of the sample. Also the probability of observing a zero-bias peak decreased monotonically with increasing magnetic field. 
These observations clearly require understanding through rigorous theoretical investigations, especially with regard to the large variation in vortex spectra, the dependence of Majorana zero-modes on magnetic field, and the effect of increasing temperature or disorder.
\\

Motivated by the experimental results, in this work we study the emergent properties of multiple Majorana bound states in vortices on the surface of FeTe$_{0.55}$Se$_{0.45}$. Following a Fu-Kane type approach~\cite{Fu2008,Marchand2012,Pikulin2017}, we describe the material system as a TI with a single  Dirac surface state that is proximity-coupled to an $s$-wave superconducting bulk. We then implant magnetic fluxes to generate disordered vortex lattices, and solve the resulting problem using the \FranzTesanovic~(FT) singular gauge transformation~\cite{Franz2000,Vafek2001}.
In doing this, we need to address certain subtleties that arise when imposing periodic boundary conditions (PBC) on finite lattices. PBC is advantageous as it allows us to dispense with edge modes and focus on low-energy states localized in vortex cores. We develop a modified FT transformation which  allows us to adapt previously established methods~\cite{Franz2000,Marinelli2000,Vafek2001,LiuFranz2015} to investigate large  vortex lattices on finite, periodic manifolds.

Using the resulting model, we first investigate the apparent discrepancies in the observation of Majorana zero-modes by simulating triangular vortex lattices with varying magnetic field strengths. The average inter-vortex distance, determined by the applied magnetic field, controls the presence or absence of a zero-bias peaks corresponding to the Majorana mode. We verify that MBSs exist at low fields for the parameters observed in the experiments but may be absent at higher fields due to increasing hybridization between nearby core states.
However a perfectly ordered, periodic vortex lattice cannot explain all experimental observations since only a fraction of vortices host Majorana zero-bias peaks in the STM measurements conducted on FeTe$_{0.55}$Se$_{0.45}$. We argue that the experimental results can be explained by considering a triangular lattice with weak disorder. Such configurational disorder typically arises as a  result of competition between intervortex interactions (which favor a periodic lattice) and spatial inhomogeneities in the underlying material which tend to pin vortices at random locations. In such a disordered vortex array translation symmetry is broken, leading to a potentially large variation in vortex spectra which is in accordance with the observations in experiments~\cite{Machida2019}.
Configurational disorder along with decreasing inter-vortex distance can explain the decrease in the fraction of vortices hosting the Majorana zero modes upon increasing magnetic field.
Finally, we show that our results can be connected to a simple disordered Majorana lattice model, where only the lowest-energy modes are considered.\\

Other theoretical works on MBSs in iron-based superconductors investigated the physics of Fe interstitial defects without external magnetic field~\cite{Jiang2019} and the effect of Zeeman coupling~\cite{Ghazaryan2020}. Majorana hybridisation and vortex disorder effects in a three-dimensional tight-binding model specific to FeTe$_{0.55}$Se$_{0.45}$ were discussed in Ref.~\cite{Chiu2020}. The latter work used open boundary conditions.
Here, we study MBS hybridization and disorder effects in a two-dimensional surface-only model using a vortex lattice on a periodic manifold. Our models notably are simpler than other detailed materials-based simulations~\cite{Chiu2020}, but nonetheless capture much of the same physical behaviour found in experiments. The local disorder effects in vortex configurations that are repeated periodically in space provide a unique perspective into the long-range physics of Majorana quasi-particles. Our approach allows the use of periodic boundary conditions and fewer on-site orbitals, such that the simulation of larger lattices without boundary effects becomes possible.
Finally, our results are in qualitative agreement with a Majorana-only  model which can simulate lattices with tens of thousands of vortices on a simple laptop computer, and might be used as input for more involved (interacting) many-body quantum simulations based on the low-energy manifold of the vortex lattice.\\

The remainder of this paper is organized as follows. In Sec.~\ref{sec:model-setup}, we first review a simple square lattice model for TI surface states proximity-coupled to an $s$-wave SC~\cite{Fu2008,Marchand2012}. We then provide details on how to construct the vortex lattice including periodic boundary conditions, using a modified version of the FT transformation~\cite{Franz2000}. In Sec.~\ref{sec:FeSC}, we investigate the resulting model and compare it with the experimental observations of MBSs in FeTe$_{0.55}$Se$_{0.45}$. We here consider the effects of both changing magnetic field strengths and vortex lattice disorder due to, e.g., inhomogeneities in the underlying material. Finally, in Sec.~\ref{sec:MajTB}, we make contact with a simple disordered vortex lattice model that involves only the lowest-energy (Majorana) modes.
Conclusions and an outlook to future work and possible interesting experiments are offered in Sec.~\ref{sec:conclusion}.

\section{Model and setup}\label{sec:model-setup}

In this section, we briefly review the model for a proximitized TI surface~\cite{Fu2008,Marchand2012,Pikulin2017} that is used throughout much of this work to describe the topological surface states of FeTe$_{0.55}$Se$_{0.45}$. We then discuss how one can impose arbitrary vortex arrays in SC systems by means of the \FranzTesanovic (FT) singular gauge transformation, including some modifications that are necessary to address finite systems with  periodic boundary conditions (PBC).

\subsection{TI-SC model Hamiltonian}\label{sec:model}

Here we briefly review the model of surface states of a strong three-dimensional topological insulator (TI) that is used throughout this work~\cite{Marchand2012}. We start by defining a simple two-dimensional square lattice model, 
\begin{equation}\label{eq:hdirac-lattice}
    h_{\textrm{0}}(\bm{k})=\lambda (\sigma^{x}\textrm{sin}k_{x}+\sigma^{y}\textrm{sin}k_{y}),
\end{equation} 
where $\sigma^{i}$ are Pauli matrices in spin space and the Fermi velocity $\sim\lambda$ re-appears as a hopping parameter in the tight-binding description below. 
With four Dirac cones in the Brillouin zone, Eq.~\eqref{eq:hdirac-lattice} cannot represent a TI which is characterised by an odd number of surface Dirac cones. This is in accord with the Nielsen-Ninomyia theorem~\cite{Nielsen1981a,Nielsen1981b}, which states that it is impossible to construct a purely 2D, TR invariant lattice model with an odd number of massless Dirac fermions. The surface states of a strong TI, with an odd number of Dirac cones, are holographic in the sense that they cannot exist without the topologically non-trivial 3D bulk. A way to resolve this issue is to introduce a mass term proportional to $\sigma_z$~\cite{Marchand2012}, parametrized as $M_{k}=m[(2-\textrm{cos}k_{x}-\textrm{cos}k_{y})-\frac{1}{4}(2-\textrm{cos}2k_{x}-\textrm{cos}2k_{y})]$. With the extra mass term, we define the Hamiltonian
\begin{equation}\label{eq:hTI}
    h_{\textrm{0}}(\bm{k})=\lambda (\sigma^{x}\textrm{sin}k_{x}+\sigma^{y}\textrm{sin}k_{y})+\sigma^{z}M_{k}-\mu,
\end{equation}
where $\mu$ is the chemical potential. The mass term gaps out all the Dirac cones except the one at $\Gamma\equiv(0,0)$.
We anticipate that adding such a TR symmetry breaking term will not compromise the physics of the TI-SC interface that we want to study because the presence of vortices will ultimately break TR symmetry anyway \cite{Pikulin2017}. Moreover the effect of TR breaking is small around the $\Gamma$ point, where $M_{k}\propto k^{4}$.\\

\begin{figure}[hbt]
    \centering
    
    \includegraphics[width=0.7\columnwidth]{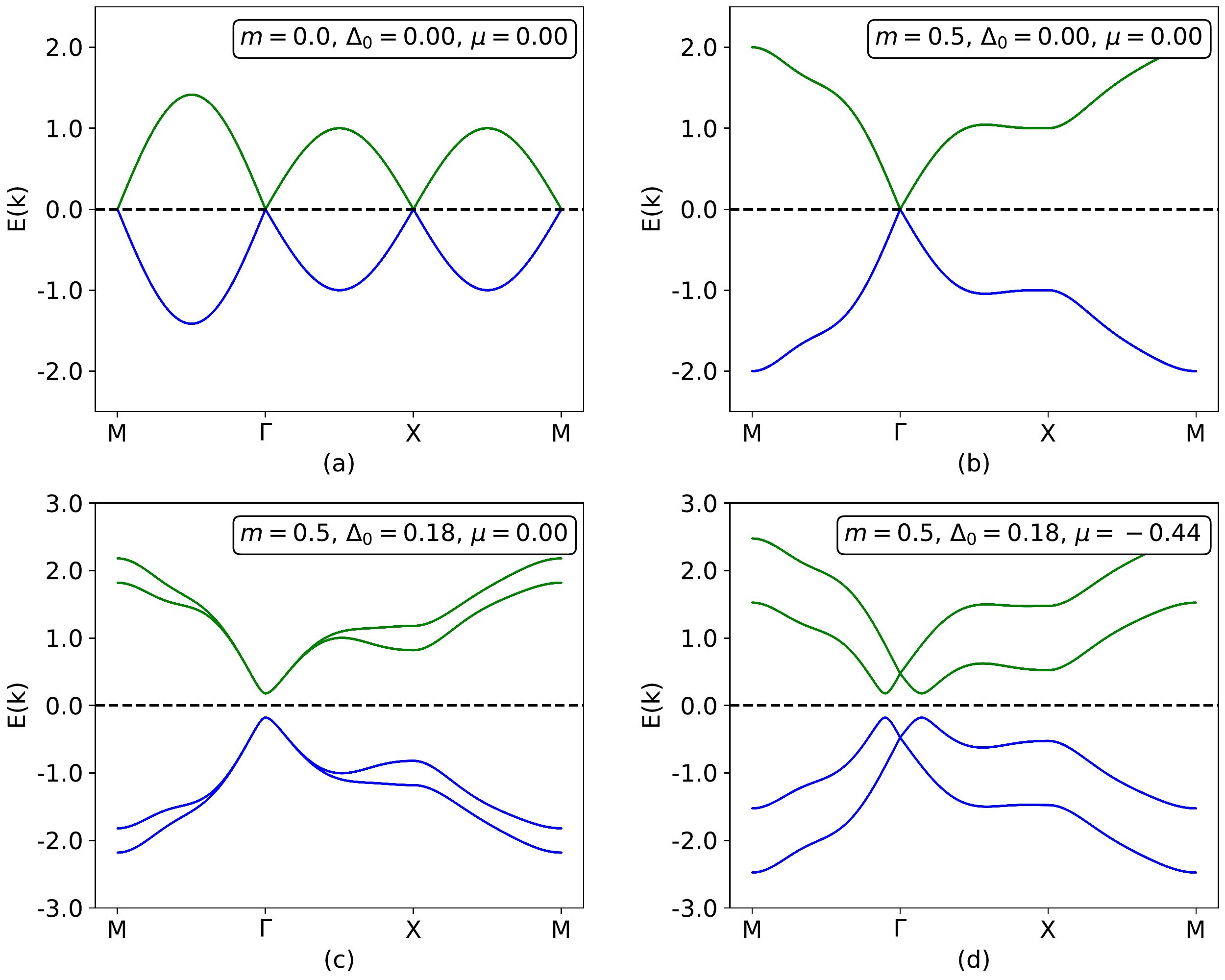}
    \caption{Band structure of the proximitized square lattice TI model in Eq.~\eqref{eq:Hk-bdg}.
    (a) without the mass term, there are four Dirac cones in the Brillouin zone. Here we show the ones at $\Gamma=(0,0)$, $X=(\pi,0)$ and $M=(\pi,\pi)$; the fourth is located at $Y=(0,\pi)$.
    (b) the mass term breaks TR symmetry and creates a gap everywhere except at $\Gamma$. The single emergent Dirac cone emulates the surface state of a 3D TI. (c) due to the proximity effect, modeled by a finite $s$-wave pairing strength $\Delta$, a superconducting gap develops in the spectrum. Spin degeneracy then is broken, except at the $\Gamma$ point. (d) same as (c), but with a finite chemical potential.}
    \label{fig:bandstructures}
\end{figure}

To incorporate the proximity-induced SC in the surface states, we use the Bogoliubov-de Gennes Hamiltonian given by
\begin{equation}
    H_{\textrm{BdG}}=\begin{pmatrix}
    h_{\textrm{0}}(\bm{k})&\Delta\\\Delta^{*}&-\sigma^{y}h_{\textrm{0}}^{*}(-\bm{k})\sigma^{y}\end{pmatrix},
\label{eq:Hk-bdg}
\end{equation}
where $\Delta$ is the induced SC order parameter which enters with the identity matrix in spin space. The term $h_{\textrm{0}}(\bm{k})=h_{\bm{k}}(\lambda,m,\mu)$ represents the particles and its time-reversed counterpart, $-\sigma^{y}h_{\textrm{0}}^{*}(-\bm{k})\sigma^{y}=-h_{\bm{k}}(\lambda,-m,\mu)$, represents the holes in this basis.
Figures~\ref{fig:bandstructures}a-b depict the spectrum of the normal state Hamiltonian in Eq.~\eqref{eq:Hk-bdg} when the mass term is zero and non-zero, respectively. Figure~\ref{fig:bandstructures}c shows the effect of a finite SC pairing $\Delta$. Figure \ref{fig:bandstructures}d depicts the bands with finite $\mu$ and $\Delta$, resulting in the formation of Dirac cones at energy $\pm\mu$ at the $\Gamma$ point. Here and in the remainder of our work we focus only on the low energy physics of the model, and choose parameters for which the non-superconducting Hamiltonian Eq.~\eqref{eq:hTI} faithfully represents a single Dirac fermion at the $\Gamma$ point.\\

Upon Fourier transforming the Hamiltonian in Eq.~\eqref{eq:hTI} leads to the second-quantized real-space lattice model for the TI surface state that reads~\cite{Pikulin2017}
\begin{eqnarray}
H_{0}&=i\frac{\lambda}{2}\sum_{\Vr,\alpha}(\psi^{\dag}_{\Vr}\sigma^{\alpha}\psi_{\Vr+\alpha}-\textrm{h.c.})+\sum_{\Vr}\psi^{\dag}_{\Vr}\left(\frac{3}{2}m\sigma^{z}-\mu\right)\psi_{\Vr}
\notag\\
&-\frac{m}{8}\sum_{\Vr,\alpha}(4\psi^{\dag}_{\Vr}\sigma^{z}\psi_{\Vr+\alpha}-\psi^{\dag}_{\Vr}\sigma^{z}\psi_{\Vr+2\alpha}+\textrm{h.c.})~.
\label{eq:pos}
\end{eqnarray}
Here $\alpha=x,~y$ labels both the hopping direction and Pauli matrices, and $\psi_{\Vr}=(c_{\Vr\uparrow},c_{\Vr\downarrow})^{T}$ denotes the two-component spinor at lattice site $\Vr$. The hopping $\lambda$, mass $m$, and chemical potential $\mu$ are as above except for an absorbed lattice constant.
Upon proximity-coupling the TI to a $s$-wave SC, Cooper pairs tunnel across their interface, inducing an effective $s$-wave pairing amplitude in the surface states of the TI. The resulting TI-SC model is written in terms of a standard BdG Hamiltonian~\cite{Pikulin2017},
\begin{equation}
\mathcal{H}=H_{0}+\sum_{\Vr}(\Delta_{\Vr}c^{\dag}_{\Vr\uparrow}c^{\dag}_{\Vr\downarrow}+\textrm{h.c.}),
\label{eq:bdgpos}
\end{equation}
where $\Delta_{\Vr}$ is the induced SC pairing at site $\Vr$. The on-site basis now expands to include both particle and hole states, and is given by the Nambu spinor
$\Psi_\Vr = (\psi_\Vr,i\sigma_y \psi_\Vr^\dagger)^T = (c_{\Vr\uparrow},c_{\Vr\downarrow},c^{\dag}_{\Vr\downarrow},-c^{\dag}_{\Vr\uparrow})^{T}$.
In this basis the pairing matrix $\Delta(\Vr)$ is proportional to the identity in spin space. Thus, we arrive at the eigenvalue problem
%
\begin{equation}\label{eq:pos_space_bdg}
    \begin{pmatrix}
    H_{\rm 0}(\lambda,\mu,m) & \Delta(\Vr) \\
    \Delta^*(\Vr) & -H_{\rm 0} (\lambda,\mu,-m)
    \end{pmatrix}
    \begin{pmatrix} \bu_n \\ \bv_n \end{pmatrix}
    = E_n \begin{pmatrix} \bu_n \\ \bv_n \end{pmatrix},
\end{equation}
where $H_{\rm 0}(\lambda,\mu,m)$ is the Hamiltonian of particles given by Eq.~\eqref{eq:pos}, and its time-reversed counterpart $-H_{\rm 0}(\lambda,\mu,-m)$ describes holes.
The spin-space vectors $\bu_{n}(\Vr)$ and $\bv_{n}(\Vr)$ encode particle and hole eigenstates, respectively, and $E_{n}$ are the corresponding energy eigenvalues. The lattice BdG Hamiltonian can likewise be viewed as stemming from a direct Fourier transform of the momentum-space version in Eq.~\eqref{eq:Hk-bdg}. Finally, let us recall that the particle-hole symmetry of the BdG Hamiltonian is a built-in constraint of the theory that must be obeyed. We discuss apparent violations of particle-hole symmetry stemming from inconsistent gauge choices, as well as their resolution, in Sec.~\ref{sec:FT} and \ref{app:A}.

\subsection{Imposing vortex lattices in systems with periodic boundary conditions}\label{sec:FT}

We now discuss how one can impose arbitrary vortex lattices in generic SC systems, within the BdG formalism and while using periodic boundary conditions (PBC).
To this end, note that for a translation-invariant system we can define a uniform superconducting order parameter, i.e. $\Delta(\Vr)=\Delta_{0}$. In the presence of vortices, however, the order parameter acquires a spatially varying phase, $\Delta(\Vr)=\Delta_{0} e^{i\phi(\Vr)}$, where $\phi(\Vr)$ winds by $2\pi$ around each vortex. Although one expects physical observables to be perfectly periodic in the presence of periodic vortex lattice the factor $e^{i\phi(\Vr)}$ lacks this periodicity thus precluding the naive inclusion of PBC. As first pointed out by Anderson~\cite{Anderson1998} and by Gorkov and Schrieffer~\cite{Gorkov1998} the situation can be remedied when the external  magnetic field is included in the theory to model a realistic, thermodynamically stable vortex lattice.
The latter is incorporated in Eq.~\eqref{eq:pos_space_bdg} via the Peierls substitution, according to which all hopping terms 
acquire a phase $\varphi_{\Vr,\Vr+\alpha} = \frac{e}{\hbar c} \int^{\Vr+\alpha}_{\Vr}\bA\cdot\bm{dr}$, with the magnetic vector potential $\bA$.
The key insight is that the lack of translation invariance in the naive BdG Hamiltonian is a gauge artifact. Correspondingly, the solution consists of finding the correct gauge in which the Hamiltonian recovers the invariance. To achieve this  we shall use a  
modified version of the FT singular gauge transformation~\cite{Franz2000,Vafek2001}, designed to circumvent the problem of branch cuts. This allows us to impose PBC, and thus to describe large vortex lattices without boundary effects.
\\

\begin{figure}[hbt]
\centering
\includegraphics[width=0.6\columnwidth]{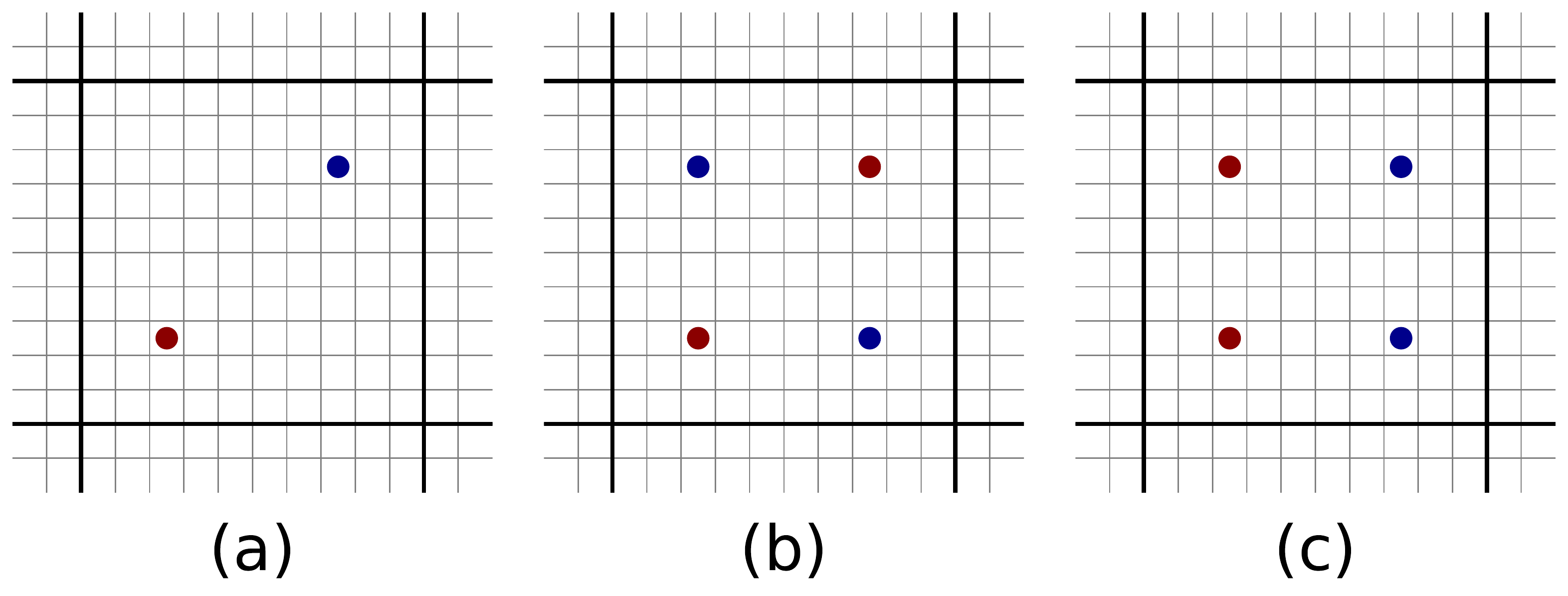}
\caption{Magnetic unit cell (black) with A (red) and B (blue) vortex sub-lattice. (a) Pattern with two square sub-lattices A and B, formed by two vortices per magnetic unit cell. (b,c) Two possible labels of the same vortex lattice dubbed ABAB (b) and AABB (c), see text, with four vortices per unit cell.}
\label{fig:ABsub-lattice}
\end{figure}

Let us consider a vortex lattice of $2n$ vortices that we arbitrarily divide into groups A and B of $n$ vortices each. Figure \ref{fig:ABsub-lattice} shows the A and B vortex sub-lattices for systems with two or four vortices per magnetic unit cell. The total phase contributions to the SC order parameter are noted as $\phi_{A}(\Vr)$ and $\phi_{B}(\Vr)$, respectively, and obey 
\begin{equation}\label{curlofgrad}
    \nabla\times\nabla\phi_{g}=2\pi\hat{z}\sum_{j}\delta(\Vr-\Vr_{g}^{j}),
\end{equation}
where $\Vr^{j}_{g}$ is the position of $j^{th}$ vortex of type $g=A,B$. With this, we can define a unitary transformation~\cite{Franz2000,Vafek2001}
\begin{equation}\label{eq:U}
    U = \begin{pmatrix} e^{i\phi_A(\Vr)} & 0 \\
    0 & e^{-i\phi_B(\Vr)} \end{pmatrix},
\end{equation}
where $\phi_A(\Vr)+\phi_B(\Vr)=\phi(\Vr)$ gives the total SC phase.
Because of the delta functions in Eq.~\eqref{curlofgrad}, the above transformation has to be regularized on the lattice~\cite{Vafek2001}.
Note that the A and B blocks in Eq.~\eqref{eq:U} refer to the Nambu structure in the BdG Hamiltonian, cf. Eq.~\eqref{eq:Hk-bdg}. Hence $U$ in Eq.~\eqref{eq:U} is not a pure gauge transformation, and the effective magnetic fields that are experienced by electrons and holes are changed.
\\

Applying the unitary Eq.~\eqref{eq:U} on the BdG Hamiltonian, $\mathcal{H}\rightarrow U^{-1}\mathcal{H}U$, removes the phase factors from the SC pairing terms in the off-diagonal blocks of Eq.~\eqref{eq:bdgpos}. The transformed BdG Hamiltonian matrix then is given by
\begin{equation}\label{newBdG}
\begin{split}
\mathcal{H} = \begin{pmatrix}
\sum_{\Vr,\alpha}h^e_{\Vr,\Vr+\alpha}\textrm{e}^{i\mathcal{V}_{A}^{\alpha}(\Vr)}&\Delta_{0}\\\Delta_0 &-\sum_{\Vr,\alpha}h^h_{\Vr,\Vr+\alpha}\textrm{e}^{-i\mathcal{V}_{B}^{\alpha}(\Vr)}
    \end{pmatrix}
\end{split}
\end{equation}
where $\mathcal{V}_{A}$ and $\mathcal{V}_{B}$ are new phase factors appearing in the hopping terms of electron and hole component. They are
\begin{equation}\label{phases}
\begin{split}
\mathcal{V}_g^\alpha (\Vr) &= \int^{\Vr+\alpha}_{\Vr}
\left(\nabla\phi_{g}(\Vr)-\frac{e}{\hslash c}\bA\right) \cdot\bm{dr}\\
&=\frac{2\pi}{\Phi_0}\int^{\Vr+\alpha}_{\Vr}
\left(\frac{\Phi_0}{2\pi} \nabla\phi_{g}(\Vr)-\bA\right) \cdot\bm{dr}
\end{split}
\end{equation}
with $g=A,B$, and $\Phi_{0}=hc/e$ the  flux quantum. The resulting total phases in Eq.~\eqref{phases} are manifestly gauge invariant, making the unitary transformation Eq.~\eqref{eq:U} a singular gauge transformation.
Further, the integrands in Eq.~\eqref{phases} can be viewed as an effective vector potential, leading to the effective magnetic field
\begin{equation} 
\begin{split}
\mathcal{\vec{B}}_{g}&=\nabla\times \left(\bA-\frac{\Phi_{0}}{2\pi}\nabla\phi_{g}(\Vr)\right)\\
&=\bm{B}-\hat{z}\cdot \Phi_0 \sum_{j}\delta(\Vr-\Vr^{j}_{g})~.
\end{split}
\end{equation}
One may interpret this as quasi-particles experiencing an effective magnetic field $\mathcal{\vec{B}}_{A}$, which comprises a uniform physical magnetic field $\vec{B}$ and delta function spikes of opposing polarity at vortices of type A. Similarly, quasi-holes experience an effective magnetic field $\mathcal{\vec{B}}_{B}$. The reason for choosing an equal number $n$ of vortices of type A and B is so that $\mathcal{\vec{B}}_{g}$ is zero when averaged over the lattice. Alternatively, there is one flux spike arising due to each vortex of type A or type B that compensates for each flux quantum of the physical magnetic field.
Now the condition $\langle\mathcal{\vec{B}}_{g}\rangle=0$ allows us to incorporate PBC in the transformed BdG Hamiltonian Eq.~\eqref{newBdG}, which had only allowed open boundary conditions prior to the application of the unitary transformation Eq.~\eqref{eq:U}. Therefore, we can create a periodic vortex lattice with arbitrarily-chosen sub-lattices of type A and B.\\

As a practical matter, it would seem that we just need to calculate the phase factors in Eq.~\eqref{phases} in a convenient manner, followed by diagonalization of the Hamiltonian Eq.~\eqref{newBdG} to find the spectrum and eigenstates.
The phase factors in Eq.~\eqref{phases} can be constructed in terms of the reciprocal lattice vectors $\bG$ of the magnetic unit cell in Fig.~\ref{fig:ABsub-lattice}, as described in  Refs.~\cite{Vafek2001,LiuFranz2015} and \ref{a1}. This results in 
\begin{equation}\label{eq:FTphases}
\mathcal{V}^\alpha_g(\Vr) = \frac{2\pi}{L_xL_y} \sum_{j, \bG} \left[\frac{i\bG\times\hat{z}}{\lambda_{s}^{-2}+G^{2}}\cdot
\int_{\Vr}^{\Vr+\alpha} e^{i\bG\cdot(\Vr-\Vr^j_g)}\bm{dr} \right],   
\end{equation}
where $\Vr^j_g$ are the vortex positions and $\lambda_{s}$ is the London penetration depth. However, there are some subtleties in applying the FT transformation in systems with PBC.

First, the application of the FT transformation must ultimately yield the same physical observables independent of the choice of A-B sub-lattice degrees of freedom. For instance, the four-vortex configurations, ABAB and AABB, illustrated in Fig. \ref{fig:ABsub-lattice}, have different assignment of A-B vortices but must be equivalent in all physical aspects. Furthermore, the particle-hole symmetry must be obeyed for any vortex configuration.
These two symmetry constraints that must be respected by any physical model are broken if one directly applies the FT recipe~\cite{Franz2000,Marinelli2000,Vafek2001} under PBC using the phase factors in Eq.~\eqref{eq:FTphases} (see \ref{sec:symmetry}).

The resolution of this apparent symmetry breaking lies in understanding that the electrons and holes on the underlying physical lattice must experience the same magnetic field. This translates into the condition that flux threaded through any plaquette of the lattice can only differ by multiples of $2\pi$, as dictated by the FT transformation. 
However, upon imposing PBC on the FT-transformed BdG Hamiltonian, we need to ensure explicitly that the fluxes through the two additional, non-contractible loops in the torus geometry (2D lattice with PBC) obey the same principle. 
This leads to a modified prescription  for the phase factor in Eq.~\eqref{eq:FTphases}, which thus are given as
\begin{equation}\label{modification1}
\mathcal{\Tilde{V}}_{A/B}^{\alpha}(\Vr) = 
\mathcal{V}_{A/B}^{\alpha}(\Vr) \pm k_{\alpha}\cdot\alpha,
\end{equation}
where $\alpha=\delta x,~\delta y$ are the hopping directions on the lattice. The correction ``wave-vector'' $\bm{k}$ here follows from
\begin{equation}\label{kxky_final1}
\bm{k}=\frac{\pi}{L}\cdot\frac{\Delta\Vr\times\hat{\bm{z}}}{L} ~,
\end{equation}
where $\Delta \Vr=\sum_{j\in A}\Vr_A^j - \sum_{j\in B}\Vr_B^j$ with $j$ running over the vortices of type A and B, respectively. A detailed derivation of this result is given in \ref{sec:revised_ft}.

\section{Majorana vortex modes in iron-based superconductors}
\label{sec:FeSC}

Here we perform simulations of vortex lattices at varying magnetic fields and disorder strength. All results are based on the TI-SC model introduced in Sec.~\ref{sec:model}, using vortex lattices under PBC that are implemented via the modified \FranzTesanovic transformation of Sec.~\ref{sec:FT}.

\subsection{Lattice setup, model parameters, and their relation to experiment}
\label{sec:setup-params}

We here show how to fit the physical triangular vortex lattice into the square lattice of the underlying materials model, cf. Sec.~\ref{sec:model}. We then discuss how one can represent the model parameters in such a way that the mean inter-vortex distance $d_v$ and the disorder strength, Sec.~\ref{sec:lattice-dis}, enter as tunable parameters. Material parameters other than the mean vortex distance (set by the applied field) are chosen fixed and correspond to sensible experimental values~\cite{Wang2018,Zhang2018,Chiu2020}, see Table~\ref{table:params}.
The mean vortex distance $d_v$ and disorder parameters, see Sec.~\ref{sec:lattice-dis}, will also be used for the simpler Majorana lattice model in Sec.~\ref{sec:MajTB}.\\

\begin{figure}[hbtp]
    \centering
    \includegraphics[width=0.6\columnwidth]{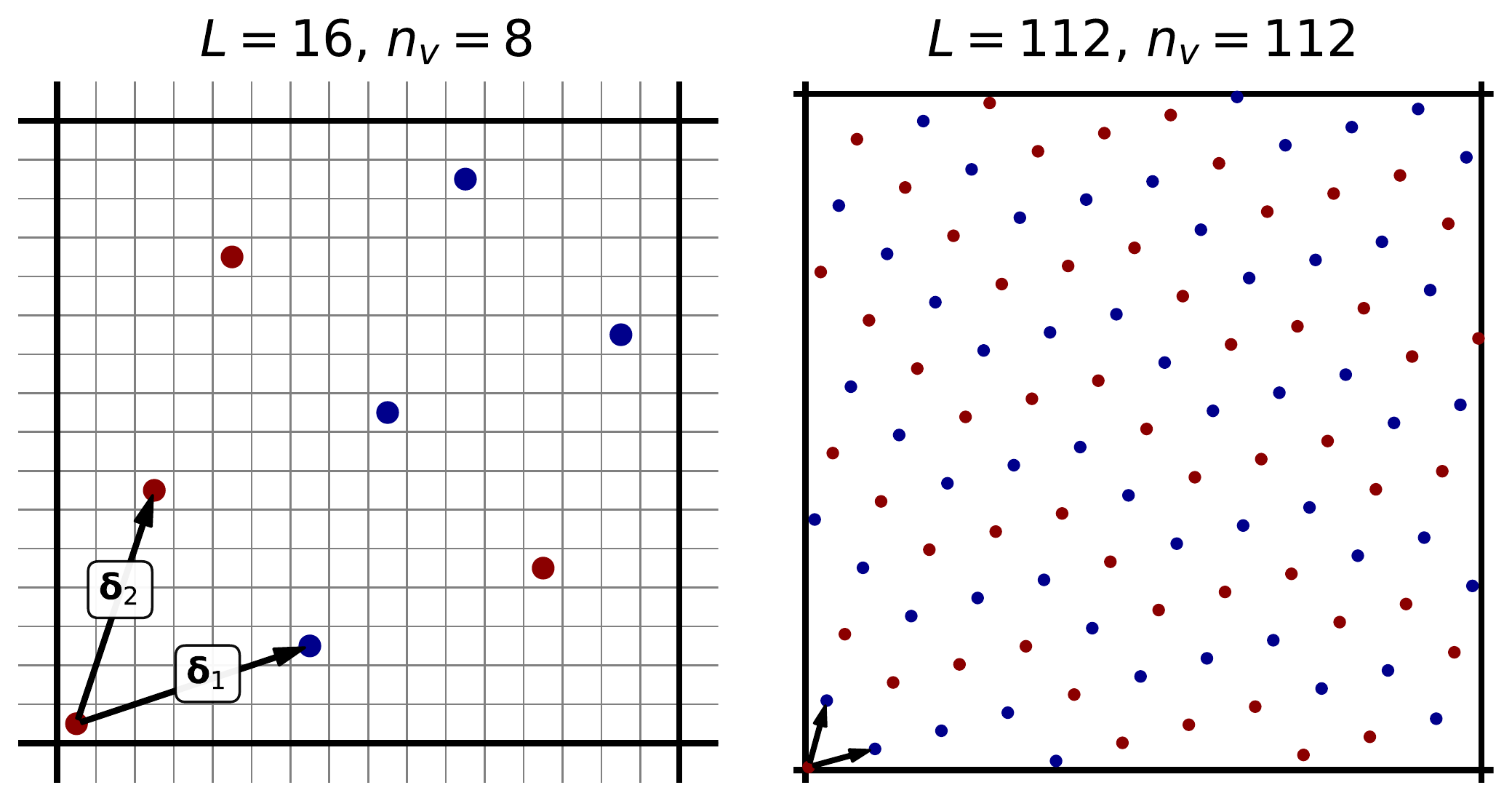}
    \caption{Setup for triangular vortex lattice simulation. Left: a magnetic unit cell of $n_v=8$ vortices in a $L=16$ lattice realizes a roughly triangular vortex lattice. Right: a larger unit cell with $n_v=112$ and $L=112$ allows to closely match the desired triangular lattice. The latter is used in our simulations, also with added disorder as discussed in Sec.~\ref{sec:lattice-dis}.}
    \label{fig:lattice_magcell}
\end{figure}

In our simulations, we choose square lattice model sizes $L_x\times L_y$ that allow to closely fit a triangular vortex pattern with the vortex positions fixed to centers of the square lattice plaquettes. Such vortex arrangements are generated from simple geometric considerations, with two example configurations depicted in Fig.~\ref{fig:lattice_magcell}. We here fix $x$ and $y$ directions to have the same number of square lattice cells, $L_{x,y}=L$, and further demand that they exhibit the same pattern of vortices, related by a mirror about the $x$-$y$ axis. This naturally keeps the two lattice directions equivalent, and defines a convenient magnetic unit cell for triangular lattice simulation.
Let us denote the two Bravais lattice vectors of the desired, ideal triangular lattice that is rotated like in Fig.~\ref{fig:lattice_magcell}, as
\begin{equation}\label{eq:basevec-def}
{\bm \delta}_1 \simeq (c_{15},s_{15})~,~~  
{\bm \delta}_2 \simeq (s_{15},c_{15})~,
\end{equation}
with $c_{15} = \cos(15^\circ)$, $s_{15} = \sin(15^\circ)$. We have omitted an overall scale factor here. The goal now is to find a lattice arrangement of vortices that allows us to approximate these vectors, but where the entries naturally are integers (viz., integer multiples of the lattice constant). Simply put, we want to find pairs of integers $(n,m)$ that approximate
\[
\frac{m}{n} \approx \frac{s_{15}}{c_{15}} = 2-\sqrt3= 0.26794\cdots,
\]
and to define lattice vectors ${\bm \delta}_1 = (n,m)$ and ${\bm \delta}_2 = (m,n)$. Integer pairs that realize gradually improving rational approximations of this number are
\begin{equation}\label{eq:nmcoeff}
(n,m) = (3,1),~(4,1),~(11,3),~(15,4),~\cdots.
\end{equation}
Furthermore, the triangular lattice generated by ${\bm \delta}_{1,2}$ has to fit into our square lattice model. Going to larger $(n,m)$ naturally demands a larger underlying lattice, which thus becomes increasingly expensive to simulate. To fix the required lattice size $L$, consider the corners $(x,y) = (0,0)$ and $(L,0)$ of the magnetic unit cell, cf.~Fig.~\ref{fig:lattice_magcell}. Since these are identified with each other, we have to be able to connect them by a sum of lattice vectors ${\bm \delta}_{1,2}$. Hence
\[ (L,~0) = k_1{\bm \delta}_1 - k_2{\bm \delta}_2 = (k_1n - k_2m,~ k_1m - k_2n) \]
with some integers $k_{1,2}$. For the pairs $(n,m)$ in Eq.~\eqref{eq:nmcoeff}, we then find coefficients
\begin{equation}
\left(\frac{L}{k_1}, \frac{L}{k_2}\right) =
\left(\frac{8}{3},\frac{8}{1}\right),~
\left(\frac{15}{4}, \frac{15}{1}\right),~
\left(\frac{112}{11},\frac{112}{3}\right),~
\left(\frac{209}{15},\frac{209}{4}\right).\qquad
\end{equation}
The possible lattice sizes hence are $L=8,~15,~112,~209,$ or multiples thereof. Once $L$ is fixed, one can read off $k_{1,2}$ above.
Finally, recall that our model necessitates an even number of vortices in the magnetic unit cell. For the above pairs $(n,m)$ and quoted lattice sizes $L$, the respective number of vortices is $n_v = 8,~15,~112,~224$. Hence the configuration $(n,m)=(4,1)$ cannot be used as magnetic unit cell, however one may repeat it to obtain a larger, admissible unit cell with - say - $L=30$ and $n_v = 60$.
In Fig.~\ref{fig:lattice_magcell} we have shown two representative examples, namely $(n,m) = (6,2)$ with $L=16$ and $n_v = 8$ and $(n,m) = (11,3)$ with $L=112$ and $n_v=112$. 
Using sparse-matrix methods for the Hamiltonian construction and diagonalization here allows us to consider large lattice sizes $L$, both to improve the resolution and to put more vortices that also more closely match a triangular lattice.
Finally, having a sizable number of vortices ($n_v=112$) will be important once we introduce positional disorder in the vortex lattice, see Sec.~\ref{sec:lattice-dis}.
%
\\

\begin{table}
\centering
\begin{tabular}{|P{2.6cm}|P{3.0cm}|P{2.6cm}|}
\hline
Parameter                       & Simulation        & Experiment    \\\hline
\hline
$\Delta_{0}$                    & 1                 & 1.8\,meV      \\\hline
$\lambda_F = k_{F}^{-1}$        & 1                 & 5.0\,nm       \\\hline
$v_{F}$                         & 2.78              & 25\,meV-nm    \\\hline
$\mu = v_F/\lambda_F$           & -2.78             & 5.0\,meV      \\\hline
$\xi= v_{F}/\Delta_0$           & 2.78              & 13.9\,nm      \\\hline
$d_v\sim B^{-1/2}$              & $\sim$ 3~-~14     & 15\,nm~-~70\,nm \\\hline
$\eta_{\rm STM}$                & $\sim 0.01$       & 20\,$\mu$ eV    \\\hline
\hline
$a_0$       & $\sim d_v$, Eq.~\eqref{eq:d-lattice}  & -             \\\hline
$\lambda=v_F/a_0$   & $\sim d_v^{-1}$, Eq.~\eqref{eq:lambda-scaled} & - \\\hline
$m$     & $\lesssim\lambda$, Sec.~\ref{sec:model}   & -             \\\hline
\end{tabular}
\caption{Simulation and experiment values of parameters, cf. Refs.~\cite{Zhang2018,Wang2018,Chiu2020}. The lattice spacing $a_0$ and model parameters $\lambda$ and $m$ are scaled such that different inter-vortex distances $d_v$ (magnetic fields $B$) can be simulated, see Eqs.~\eqref{eq:d-lattice}, \eqref{eq:lambda-scaled} and discussion in text. The mass $m\simeq 0.5\lambda-1\lambda$ is not fixed by a strict rule, but rather chosen to ensure a sizable gap at all points in the Brillouin zone away from $\Gamma=(0,0)$, cf. Fig.~\ref{fig:bandstructures}.}
\label{table:params}
\end{table}

After fixing the lattice setup we now discuss parameter choices and their relation to experiments, including how one can simulate different mean inter-vortex distances $d_v$ in fixed-size lattices by rescaling.
First, the experimentally determined parameters in
the FeTe$_{0.55}$Se$_{0.45}$ materials platform are the SC gap $\Delta_0$, the Fermi velocity $v_F$, and the Fermi wave-vector $k_F$. Their values are quoted in Table~\ref{table:params}. In our simulations, we choose the SC gap $\Delta_0$ as the unit of energy, and the Fermi wave-length $\lambda_F = k_F^{-1}$ as basic length scale. From these, the chemical potential $\mu$ and Majorana coherence length $\xi$ follow via well-known relations for the model in Sec.~\ref{sec:model}. Additional externally set parameters are the inter-vortex distance $d_v$ and the STM resolution $\eta_{\rm STM}$, see Table~\ref{table:params}. For the triangular lattice, the inter-vortex distance follows by simple geometric consideration as $d_v = (2\Phi_0/\sqrt{3}B)^{1/2}$, with the magnetic flux quantum $\Phi_0$ and applied magnetic field $B$.
We take typical values $B=1$T$-8$T, i.e. $d_v=70$nm$-17$nm.
Now for $n_v$ vortices in a lattice model of size $L\times L$, the vortex distance is given by $d_v = d_v^{\rm lattice} a_0$, with
\begin{equation}\label{eq:d-lattice}
d_v^{\rm lattice} = \left(\frac{2}{\sqrt{3}n_v}\right)^{1/2} L
\end{equation}
the vortex distance in lattice units. Since we want to set $d_v$ as an external parameter while $d_v^{\rm lattice}$ is already fixed by the lattice setup $(L,n_v)$, we will thus have to re-scale the lattice constant $a_0$ in our model. %
The scaled lattice constant $a_0$ also translates to rescaled model parameters $\lambda$ and $m$. In particular, the hopping parameter $\lambda$ is determined via Fermi velocity and lattice spacing as
\begin{equation}\label{eq:lambda-scaled}
\lambda = \frac{v_F}{a_0} \simeq 2.78\Delta_0\lambda_F \cdot d_v^{\rm lattice}/d_v~,
\end{equation}
where we inserted natural units, cf. Table~\ref{table:params}.
The mass $m$ then is chosen as $m\simeq 0.5\lambda-1.0\lambda$, cf. Sec.~\ref{sec:model}.
Equivalent arguments apply to the square lattice, where for the $n_v=2$ unit cell in Fig.~\ref{fig:ABsub-lattice} one has $d_v^{\rm lattice} = L/\sqrt2$.\\

It may seem unconventional that the lattice spacing $a_0$ and model parameters $\lambda$ and $m$ are varied together with the vortex distance $d_v$, and hence with the applied magnetic field $B$. To this end, it helps to remember that the only crucial feature of the TI surface state model in Sec.~\ref{sec:model} was the presence of an isolated Dirac cone at the $\Gamma$ point. This fact is not altered by the above parameter scaling procedure. The square crystal lattice should be viewed as a means to regularize the model at short distances and thus allow for straightforward numerical simulation.  An intuitive physical picture is that our simulation keeps the number of vortices $n_v$ in the investigated ``field of view'' constant, i.e. while increasing the magnetic field we also zoom in on a smaller patch of the material. As discussed above, this approach allows us to consider a fixed number of vortices for which the triangular vortex lattice fits well into the fixed-size materials model.

\subsection{LDOS, spatial DOS, and MVM hybridization}
\label{sec:LDOSdef}

In figures below, cf.~Fig.~\ref{fig:MVMhyb_2v}, we show the local density of states (LDOS) integrated over a window around a vortex position $\Vr_v$, and the spatially resolved DOS at zero energy or at energies that correspond to peaks in the LDOS. To make this concrete, let us denote the position and energy-resolved DOS as $\rho(\Vr,\omega)$. For site $\Vr$ of the lattice and for each eigenstate $j$ of the Hamiltonian, we have a four-component spin-Nambu vector $\Psi_j(\Vr) = [\bu_j(\Vr), \bv_j(\Vr)]^T$ with electron- and hole-components $\bu_j(\Vr)$ and $\bv_j(\Vr)$.
The spatially and frequency-resolved DOS then follows as
\begin{equation}\label{eq:LDOSdef-full}
\rho(\Vr,\omega) = \sum_{j:E_j>0} \left\lbrace |\bu_j(\Vr)|^2 \delta(\omega-E_j) + |\bv_j(\Vr)|^2 \delta(\omega+E_j) \right\rbrace.
\end{equation}
Obviously this quantity contains too much information to show in a simple plot. Instead, the LDOS at a vortex position $\Vr_v$ can be calculated as
\begin{equation}\label{eq:LDOSdef}
{\rm LDOS}_{\Vr_{v}}(\omega) = \sum_{|x-x_v|<W}\sum_{|y-y_v|<W}\rho((x,y),\omega)~,
\end{equation}
with integration window of size $2W\times 2W$. The spatially resolved DOS at energy $E$ is ${\rm DOS}_E(\Vr) = \rho(\Vr,\omega=E)$, where we omit the integration over an energy window. Rather the spectral broadening $\eta$ (STM resolution $\eta_{\rm STM}$) is included in the calculation of spectral densities $\rho(\Vr,\omega)$, where we put Lorentzian line shapes of width $\eta$. Alternatively, one could also consider a thermally broadened peak at temperature $T$, i.e. $\delta_T(\omega) \sim 1/\cosh^2(\omega/T)$.
Note that the LDOS as defined here is not necessarily particle-hole symmetric. While one may start with a PH symmetric $\rho(\Vr,\omega)$ in Eq.~\eqref{eq:LDOSdef-full}, the quantity in Eq.~\eqref{eq:LDOSdef} reflects what is measured as LDOS in experiments.
Indeed, we find that the LDOS plots shown in Fig.~\ref{fig:MVMhyb_2v} and Secs.~\ref{sec:TISC_triag} and~\ref{sec:TISC_triag_dis} generally are PH asymmetric even for non-disordered vortex lattices, though the effects are more pronounced when disorder is included. This happens even though the total, spatially averaged DOS is necessarily PH symmetric, and indicates that pairs of particle and hole eigenstates can have spatially distinct spectral weight distributions. By picking a fixed-size spatial window to represent the ``vortex LDOS'', we then implicitly choose how much spectral weight of individual eigenstates is included.
\\

\begin{figure}[hbtp]
    \centering
    \includegraphics[width=0.7\columnwidth]{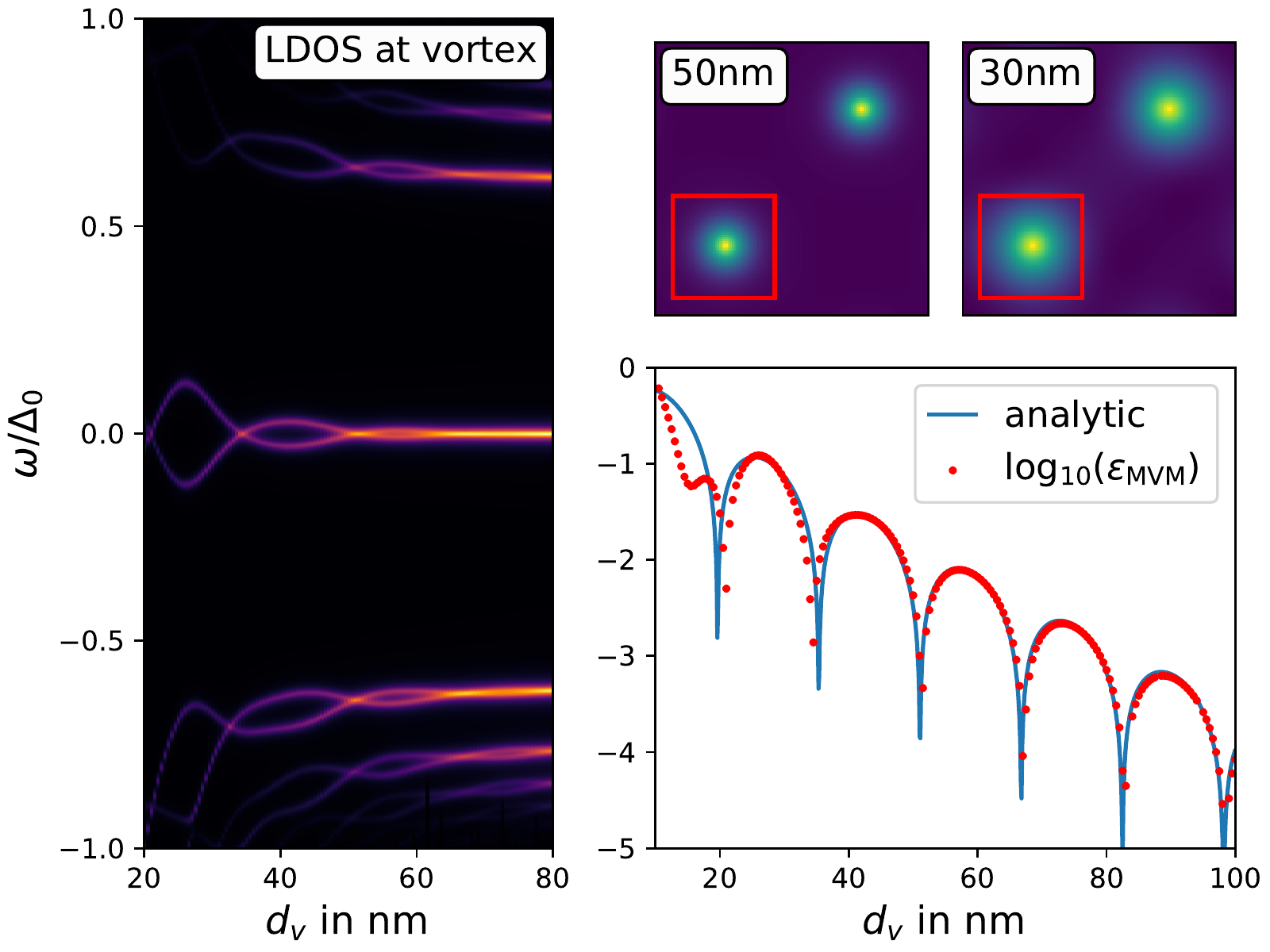}
    \caption{Local density of states (LDOS) at the vortex, and MVM hybridization $\varepsilon_{\rm MVM}$, vs vortex separation $d_v$. We here consider an ordered square vortex lattice with two vortices per unit cell, cf.~Fig.~\ref{fig:ABsub-lattice}(a), with $L=128$. The top right plots show the spatially resolved DOS at zero energy, for vortex separations $d_v = 30\,{\rm nm},~50\,{\rm nm}$. The red boxes indicate the area that we integrate over to obtain the LDOS (left plot). The bottom right plot shows the numerically obtained energy splitting that closely matches the analytical expectation for the MVM hybridization (blue), cf. Eq.~\eqref{eq:MVMhyb}.
    The lattice is kept invariant, and the effective vortex distance $d_v$ is changed by scaling the lattice constant, see Eq.~\eqref{eq:lambda-scaled} and discussion.}
    \label{fig:MVMhyb_2v}
\end{figure}

To check that the scaled model introduced in Secs.~\ref{sec:model} and \ref{sec:setup-params} correctly reproduces the vortex physics of proximitized TI surface states, we simulate the simplest square lattice with two-vortex unit cell, cf. Fig.~\ref{fig:ABsub-lattice}(a). The Majorana vortex mode (MVM) hybridization is expected to follow the analytical prediction~\cite{Cheng2009}
\begin{equation}\label{eq:MVMhyb}
\varepsilon_{\rm MVM}(d_v) = t_0\frac{\cos(k_F d_v +\theta)}{\sqrt{d_v}} e^{-d_v/\xi}~,
\end{equation}
with Fermi momentum $k_F$, phase $\theta$, and coherence length $\xi$. Following Refs.~\cite{Chiu2020,Cheng2009} and Table~\ref{table:params}, we take $k_F^{-1} = 5{\rm nm}$, $\theta = \pi/4$ (or $\theta \approx 1.4$), and $\xi = 13.9{\rm nm}$.
As shown in Fig.~\ref{fig:MVMhyb_2v}, our model indeed closely matches the analytical formula for any $d_v \gtrsim 20\,{\rm nm}$. We here set $t_0 = 4\Delta_0$ and $\theta = \pi/4$ to fit our numerical results. For smaller vortex separations, the hybridization of MVMs and CdGM modes becomes so strong that higher-energy features of our model become important. These are non-universal, i.e. they do not reflect an isolated TI surface Dirac cone, and hence this breakdown at very high vortex densities (magnetic fields) is expected.
Going forward, we thus are confident that our model faithfully reproduces the low-energy physics even for triangular or disordered vortex lattices, so long as the typical vortex separations are sufficiently large, $d_v^{\rm typ} \gtrsim 20\,{\rm nm}$.

\subsection{Results for the regular triangular vortex lattice}
\label{sec:TISC_triag}

\begin{figure*}[hbtp]
    \centering
    \includegraphics[width=\textwidth]{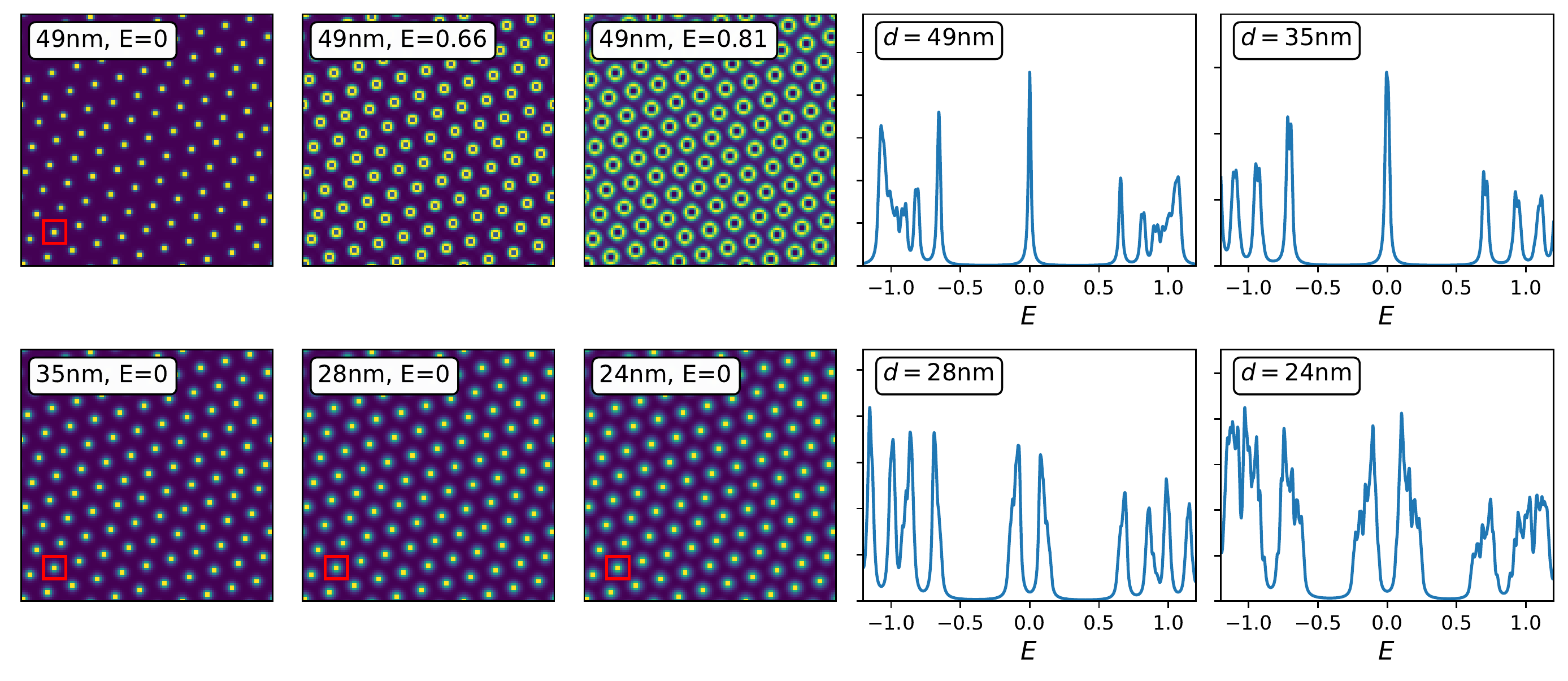}
    \caption{
    Spatial DOS at fixed energies (left) and frequency-resolved LDOS at vortex positions (right), for the triangular lattice with $L=112$ and $n_v=112$ vortices, cf. Fig.~\ref{fig:lattice_magcell}, and vortex separations $d_v \approx 49{\rm nm},~35{\rm nm},~28{\rm nm},~24{\rm nm}$ [$B= 1{\rm T},~2{\rm T},~3{\rm T},~4{\rm T}$].
    The top three plots (left) show the DOS for $d_v \approx 49 {\rm nm}$ and at energies corresponding to the first three peaks in LDOS (right). The latter is obtained by integrating the full DOS over the area of the red box in the DOS plot, cf. Sec.~\ref{sec:LDOSdef} and Eq.~\eqref{eq:LDOSdef}.
    For $d_v \approx 49 {\rm nm}$, vortices are well-separated and the spectral peaks of zero-energy and CdGM modes in the LDOS are sharp. In the DOS, CdGM modes are more spread out, and the second CdGM mode indeed forms a ring around the vortex center.
    Decreasing the vortex distance $d_v$ then leads the CdGM modes and also the lowest-energy modes to overlap (DOS) and split energetically (LDOS). For $d_v \approx 35 {\rm nm}$, only the less localized CdGM modes overlap and split. For $d_v \approx 28 {\rm nm}$, also the lowest-energy modes of distinct vortices overlap, leading to significant energy splittings.
    At large fields, $d_v \approx 24 {\rm nm}$, the strongly hybridized higher-energy modes cannot be identified as individual modes anymore. Also the low-energy modes split strongly and develop multi-peak structures. Further data for the LDOS and peak splitting under varying vortex separation is shown in Fig.~\ref{fig:MVMhyb_triag}.
    }
    \label{fig:LDOS_overview}
\end{figure*}

We now investigate the triangular vortex lattice, using the setup of $L=112$ and $n_v=112$ vortices laid out in Sec.~\ref{sec:setup-params} and Fig.~\ref{fig:lattice_magcell}. Vortex lattice disorder will be introduced and simulated in Secs.~\ref{sec:lattice-dis} and \ref{sec:TISC_triag_dis}.
Before repeating the comprehensive analysis of Fig.~\ref{fig:MVMhyb_2v} for the triangular lattice, let us consider a few typical vortex separations $d_v \approx 49\,{\rm nm},~35\,{\rm nm},~28\,{\rm nm},~24\,{\rm nm}$ (magnetic fields $B= 1\,{\rm T},~2\,{\rm T},~3\,{\rm T},~4\,{\rm T}$) and review the simulation results in more detail.
Results for both the spatially-resolved DOS at various energies and the frequency-resolved LDOS at a vortex position are shown in Fig.~\ref{fig:LDOS_overview}. For vortex separation $d_v \approx 49\,{\rm nm}$ we clearly recover the expected low-energy Majorana vortex modes and higher-energy CdGM modes, both resolved in real space (DOS) and in the local vortex spectrum (LDOS). In fact, at such a large vortex separation $d_v$, the spectral peaks are relatively sharp and not split in energy. From Sec.~\ref{sec:LDOSdef} and Fig.~\ref{fig:MVMhyb_2v}, recall that the MVM hybridization and thus the low-energy peak splitting closely followed the analytical prediction in Eq.~\eqref{eq:MVMhyb}. For the above quoted vortex separations $d_v$, we would thus obtain
\begin{equation*}
|\varepsilon_{\rm MVM}(d_v)| = 0.007\Delta_0,~0.009\Delta_0,~0.098\Delta_0,~0.114\Delta_0.
\end{equation*}
In the calculation of spectral densities we set the broadening $\eta=0.005\Delta_0$, roughly reflecting the experimental STM resolution, cf. Table~\ref{table:params}. This explains why for $d_v \approx 49\,{\rm nm}$ and $35\,{\rm nm}$ the MVM hybridization and splitting is not yet resolved. However note that higher-energy modes generally show a stronger hybridization and splitting, simply because they are less strongly localized and thus overlap even for larger vortex separations. This can clearly be observed in the LDOS for the second CdGM mode (third spectral peak) for $d_v \approx 49\,{\rm nm}$, and for both first and second CdGM modes (second and third spectral peaks) at $d_v \approx 35\,{\rm nm}$, see Fig.~\ref{fig:LDOS_overview}.\\

\begin{figure}[hbtp]
    \centering
    \includegraphics[width=0.7\columnwidth]{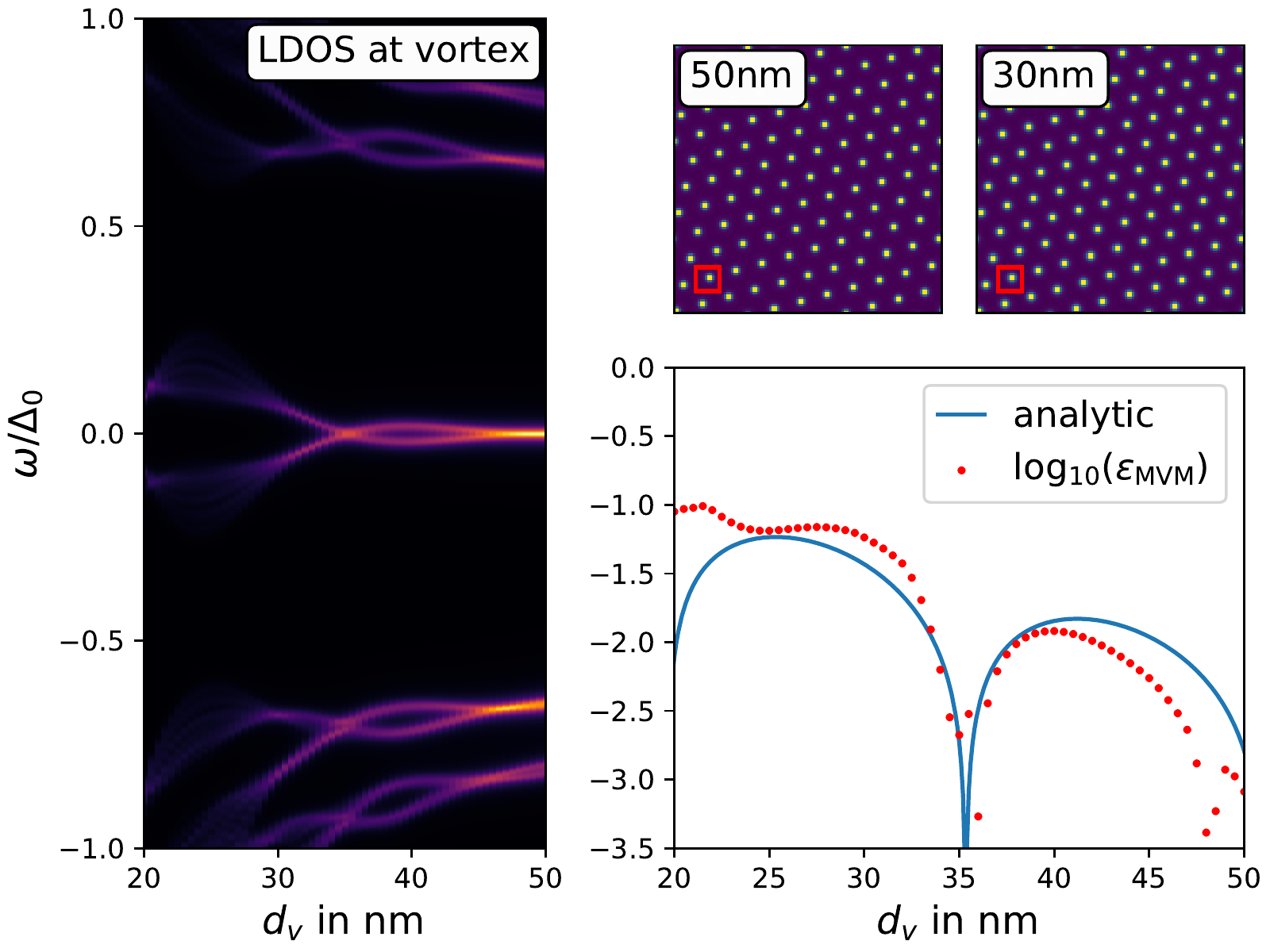}
    \caption{
    LDOS at a vortex and MVM hybridization $\varepsilon_{\rm MVM}$ vs vortex separation $d_v$, for the $L=112$ and $n_v=112$ vortex lattice.
    Top right plots show the spatial DOS at zero energy for $d_v = 50\,{\rm nm},~30\,{\rm nm}$ $[B \approx 1\,{\rm T},~3\,{\rm T}]$.
    With the spectral broadening $\eta=0.005\Delta_0$, multi-peak structures become visible in the LDOS for vortex separations $d_v \lesssim 30\,{\rm nm}$, cf. Fig.~\ref{fig:LDOS_overview}.
    The numerically obtained energy splitting of the low-energy MVM modes (bottom right) shows deviations from the naive analytical expectation in Eq.~\eqref{eq:MVMhyb}, also compared to Fig.~\ref{fig:MVMhyb_2v}.
    }
    \label{fig:MVMhyb_triag}
\end{figure}

At higher magnetic fields, for $d_v \approx 28\,{\rm nm}$ and $24\,{\rm nm}$, it generally becomes harder to distinguish individual CdGM modes. Due to their strong spatial overlap and hybridization, the CdGM modes here form broad bands rather than individual localized vortex modes. Also the lowest-energy MVMs become split, reflecting the above sizable MVM hybridization $\sim 0.1\Delta_0$, and eventually broaden into bands with a number of higher-energy but lower-intensity side peaks. This is in contrast to what we found for the simple square vortex lattice in Fig.~\ref{fig:MVMhyb_2v}, where the vortex LDOS showed a sharp pair of low-energy peaks (split by the MVM hybridization) down to small vortex separations $d_v = 20\,{\rm nm}$. 
To better understand this, we now analyze the LDOS and peak splitting behavior for vortex separations $d_v = 20{\rm nm},..., 50{\rm nm}$, see Fig.~\ref{fig:MVMhyb_triag}.
\\

The LDOS at a vortex and the MVM hybridization, for the triangular lattice with varying vortex separation $d_v$, are shown in Fig.~\ref{fig:MVMhyb_triag}. While the overall qualitative behavior of the vortex LDOS is similar as for the square vortex lattice, cf. Fig.~\ref{fig:MVMhyb_2v}, it differs in important details. Markedly, the MVM energy splittings do not quantitatively follow the analytical prediction~\cite{Cheng2009} in Eq.~\eqref{eq:MVMhyb} that was obeyed in the square lattice case. Further, as already pointed out in Fig.~\ref{fig:LDOS_overview}, both high-energy CdGM modes but also the low-energy Majorana vortex modes for small vortex separations $d_v \lesssim 30\,{\rm nm}$ develop multi-peak structures with several lower-intensity ``side bands''. It thus may appear that the low-energy physics of our model at high vortex density (large field) should not be interpreted in terms of a simple triangular Majorana lattice.
\\

We here identify two main reasons for the breakdown of the naive analytical picture that would predict MVM peak splittings with a gap given by Eq.~\eqref{eq:MVMhyb}.
First, the Majorana hybridization formula Eq.~\eqref{eq:MVMhyb} was derived for the case of two vortices at fixed distance $d$~\cite{Cheng2009}. In that case, both the tunnel coupling and thus the energy splitting of the MVMs is given by $\varepsilon_{\rm MVM}$. In a triangular lattice, however, even if one directly simulates a Majorana-only model with tunnel couplings given by Eq.~\eqref{eq:MVMhyb}, cf. Sec.~\ref{sec:MajTB} below, one does not find a single ``Majorana band'' but rather a band with multiple side-peaks~\cite{KrausStern2011,Laumann2012}. Roughly speaking, this is because it is not possible to define local pairs of MVMs that hybridize and split in a triangular lattice, because the latter is frustrated. This is in contrast with the square vortex lattice, cf. Fig.~\ref{fig:MVMhyb_2v}, which is not frustrated and where the observed peak splitting closely follows the prediction of Eq.~\eqref{eq:MVMhyb}. Having multiple side-peaks for energetically split MVMs thus is not a flaw of our model, but rather a feature that correctly reproduces the expected triangular Majorana lattice physics.

Second, as introduced in Sec.~\ref{sec:setup-params}, the triangular vortex arrays are embedded in a microscopic model that lives on a square lattice. Hence, there will always be small deviations from an ideal Majorana-only model, cf. Sec.~\ref{sec:MajTB}, for which one can freely choose the arrangement of vortices. Since the MVM tunnel couplings are strongly dependent on distance, this effect should not be underestimated. To be more quantitative, let us revisit the embedded vortex lattice construction in Sec.~\ref{sec:setup-params}. For the $L=112$ and $n_v=112$ vortex case, the lattice vectors were ${\bm \delta}_1 = (11,3)$ and ${\bm \delta}_2 = (3,11)$ (in units $a_0$). The base triangle spanning the lattice in Fig.~\ref{fig:lattice_magcell} thus has distinct side lengths $|{\bm \delta}_1| = |{\bm \delta}_2| \approx 11.4$, and $|{\bm \delta}_1 - {\bm \delta}_2| \approx 11.3$. Entering this into the MVM hybridization formula~\eqref{eq:MVMhyb}, one finds couplings $t_{\rm long}$ and $t_{\rm short}$ with the ratio
\begin{equation*}
\frac{t_{\rm long}}{t_{\rm short}} \approx 0.83,~0.71,~0.97,~1.01,
\end{equation*}
for vortex separations $d_v \approx 49\,{\rm nm},~35\,{\rm nm},~28\,{\rm nm},~24\,{\rm nm}$, respectively.
Even if we directly simulate the Majorana-only lattice model for the near-triangular lattices used in this section, we thus expect fairly significant deviations from the perfect triangular lattice case. While from a technical standpoint the results our model produces are correct, they do not quite reflect the triangular lattice configurations we were after. In the Majorana lattice model below, see Sec.~\ref{sec:MajTB}, we thus choose to work based on ideal (non-embedded) triangular lattices.
To validate our above assessment, we simulated the ideal triangular Majorana lattice model for a small number of modes $(n_v=120)$, and extracted the low-energy vortex LDOS and spectral gap. We here recovered the ``fanning out'' of the lowest-energy peak as in Fig.~\ref{fig:MVMhyb_triag}, while the gap more closely traces the analytical result in Eq.~\eqref{eq:MVMhyb}, with $t_0 = 2\Delta_0$ and $\theta = \pi/4$. (Note a factor two in the gap energy scale compared to the square lattice, cf. Sec.~\ref{sec:LDOSdef} and Fig.~\ref{fig:MVMhyb_2v}.)
\\

We now compare our results with experimental observations~\cite{Wang2018,Machida2019,Kong2019,Kong2020,Chen2018}. Due to translational symmetry, a perfect triangular lattice shows a uniform tunneling spectrum (LDOS) for all vortices, which does not happen in experiment. Setting aside minor deviations from the perfect triangular lattice, our results also show that the splitting of low-energy peaks in the LDOS should exhibit periodic oscillations, cf. Fig.~\ref{fig:MVMhyb_triag}.
While the general trend of obtaining fewer vortices with zero-bias peaks for higher fields (smaller vortex separations) is consistent with a growing Majorana hybridization, oscillations were never seen~\cite{Wang2018,Machida2019,Kong2019,Kong2020,Chen2018}.
Lastly, we observed a fanning out of the low-energy spectral peaks which gives a cluster of side-peaks with lower spectral weight toward higher energies. This might be a first indicator for the development of ``pyramid-shape'' distributions in the LDOS peak statistics~\cite{Machida2019,Chiu2020}, which we will discuss in detail for the disordered vortex lattice.

As seen here, hybridization effects alone cannot explain all experimental observations~\cite{Machida2019,Chiu2020}. In the simulations up till now, vortices necessarily are identical due to translational invariance. Instead in actual experiments there are prominent variations between the individual vortex spectra~\cite{Machida2019}, implying that the translational symmetry of the vortex lattice is broken and disorder effects need to be taken into account.
These may include positional disorder (vortices shifted from regular lattice sites), as well as fluctuations of various parameters such as the SC pairing amplitude and chemical potential. We note that Refs.~\cite{Machida2019,Chiu2020} have explored various possible sources of disorder, and concluded that most likely positional disorder is the culprit behind the strong vortex-to-vortex variations. We thus focus on the latter mechanism, and compare our results to their conclusions where appropriate.

\subsection{Generating disordered vortex configurations}
\label{sec:lattice-dis}

For the simulation of disordered vortex lattices, a key ingredient is the generation of ``good'' vortex lattice configurations that qualitatively agree with those observed in experiment~\cite{Machida2019,Chiu2020}. Once generated, we can easily compare and qualify them by inspecting  the reciprocal space structure function $S(\bm{k})=N_v^{-1}\sum_{j,k}e^{-i\bm{k}\cdot(\Vr_j-\Vr_k)}$ or by computing pair-correlation functions.
Let us denote the vortex positions in the disordered lattice as
\begin{equation}\label{eq:vortexpos}
\Vr_j = \VbR_j + \bm{w}_j~,
\end{equation}
with $\VbR_j$ the near-triangular Bravais lattice sites, see the discussion in Sec.~\ref{sec:setup-params} and Fig.~\ref{fig:lattice_magcell}, and $\bm{w}_j$ the deviations.
\\
Inspecting the experimental data in Ref.~\cite{Machida2019}, cf.~Ref.~\cite{Chiu2020}, the vortices appear to be disordered globally while locally repellent and in a roughly triangular arrangement. This is in agreement with the basic understanding of vortex physics in superconductors. As an input to the initialization of $\bm{w}_j$, we hence consider two parameters: the ``displacement variance'' $\sigma_d$, and ``disorder correlation length'' $\lambda_d$. Here $\sigma_d$ signifies the typical allowed displacement of each vortex from its triangular lattice position, independent of its neighbors, while $\lambda_d$ describes the typical length scale on which vortices tend to be displaced in a similar way - thus locally retaining a triangular lattice arrangement.
\\

An operational means by which we initialize the displacements $\bm{w}_j$, with the above prescription for a lattice of $n_v$ vortices, is as follows.
Consider the set of all $\bm{w}_j(\Vr)$ as $n_v$ separate Gaussian random variables with a fixed covariance matrix $K$. The covariance matrix can be specified by the expectation values of products of pairs of different $\bm{w}_j$, and for cross-correlated disorder (as we want to consider here) will have finite off-diagonal elements.
The displacements $\bm{w}_j$ then are chosen as Gaussian random variables with zero mean, $\langle\bm{w_j}\rangle = 0$, and (co)variance
\begin{equation}\label{eq:covmat}
K_{jk} = \langle w_j^x  w_k^x\rangle = \langle w_j^y  w_k^y\rangle = \sigma_d^2 e^{-R_{jk}/\lambda_d}~.
\end{equation}
For simplicity we here take independent displacements in $x$ and $y$ directions with identical covariance matrix, i.e. $K^{xx} = K^{yy} = K$ and $K^{xy}_{jk} = \langle w_j^x w_k^y\rangle = 0$. Note that the separations $R_{jk} = |\VbR_j - \VbR_k|$ entering Eq.~\eqref{eq:covmat} refer to the Bravais lattice sites $\VbR_j$ in Eq.~\eqref{eq:vortexpos} and Fig.~\ref{fig:lattice_magcell}.
To generate random samples of displacements $\bm{w}_j$, one thus has to specify the covariance matrix $K$ for the desired vortex lattice (with fixed $n_v$ and $L$) and for the pair of parameters $\sigma_d$ and $\lambda_d$. Independent randomly displaced vortices are obtained for $\lambda_d \to 0$, while values larger than the typical inter-vortex separation ($\lambda_d \gg d$) lead to ``quasi-ordered'' lattices.

\begin{figure}[hbpt]
	\centering
	\includegraphics[width=0.6\columnwidth]{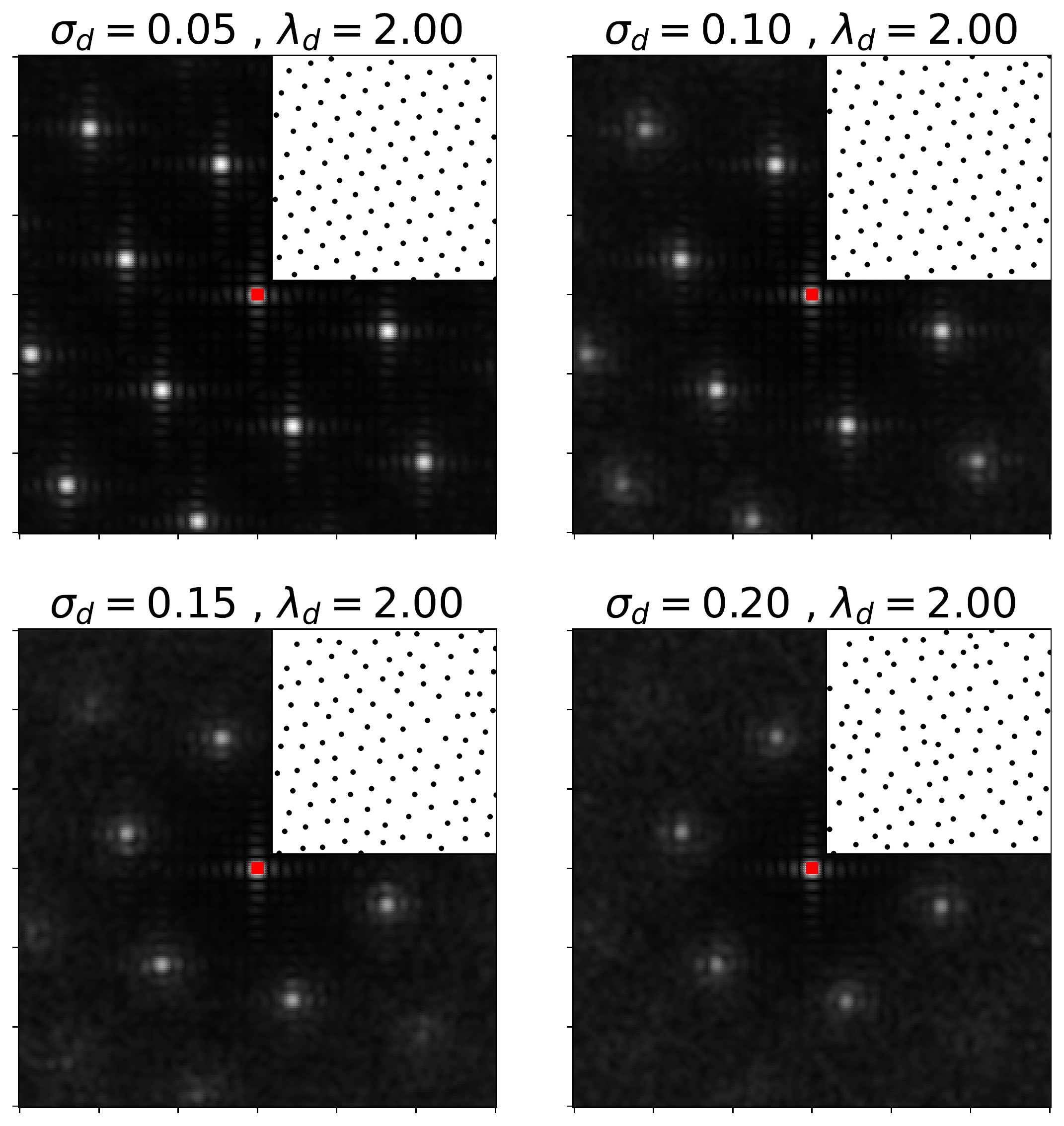}
	\caption{
	Disordered vortex lattice (inset) and the corresponding structure function $S(\bm{k})$ for varying $\sigma_d$ and $\lambda_d=2$, cf. Eq.~\eqref{eq:covmat}. Reciprocal space plots are averaged over $10$ disorder realizations. Increasing the variance $\sigma_d$ quickly diminishes distinct peaks in reciprocal space. A finite $\lambda_d$ counteracts this tendency for the peaks closest to the origin in reciprocal space (red dot), since a local ordering of vortices is preserved.
	By tuning both parameters, one can produce lattices and reciprocal-space pictures that closely mimic those seen in experiment, cf. Refs.~\cite{Machida2019,Chiu2020}.
	Note the horizontal and vertical ripples emanating from peaks in reciprocal space, indicating residual effects due to the embedding of vortices into the materials model and due to the use of periodic boundary conditions, see the discussion in Secs.~\ref{sec:setup-params} and \ref{sec:lattice-dis}.
	}
	\label{fig:lattice_dis}
\end{figure}
From the above prescription we generated a few sample lattices with different $\sigma_d$ and $\lambda_d$, and inspected the corresponding structure functions as done in Ref.~\cite{Machida2019}. Since overall scales do not matter at this moment, we set the vortex lattice spacing $d_v = 1$. We found that parameter ranges $\sigma_d \in [0, 0.25]$ and $\lambda_d \in [0, 3]$ are reasonable for reproducing experiment-like vortex configurations~\cite{Machida2019,Chiu2020}. Here one may increase $\lambda_d$ as $\sigma_d$ grows larger, such that a local approximate triangular lattice is retained.
In our disordered vortex lattice simulations below, we fix $\lambda_d = 2$ and $\sigma_d \in [0, 0.25]$. The corresponding real- and reciprocal-space lattices are shown in Fig.~\ref{fig:lattice_dis}.

We now have to address one additional issue, which already affected the setup of vortex lattices in Sec.~\ref{sec:setup-params}.
Below we want to simulate disordered vortex lattices in the FeTe$_{0.55}$Se$_{0.45}$ platform, extending the results of Sec.~\ref{sec:TISC_triag}. As with the embedding of regular vortex lattices, cf. Sec.~\ref{sec:setup-params}, we here have to amend the disordered vortex configurations such that they fit into the underlying square lattice of the materials model.
Let us start with the non-disordered but slightly distorted triangular lattice in Fig.~\ref{fig:lattice_magcell}, where we fixed vortices to sit in the centers of the materials lattice plaquettes. Now upon adding disorder, the vortex positions are displaced from the lattice plaquette centers, or even shifted to neighboring plaquettes. We then need to make sure that vortices keep a ``safe'' minimum distance from materials lattice sites, since otherwise we have to include more and more reciprocal lattice vectors in the calculation of FT phase factors. This is because the cutoff for sums over reciprocal lattice vectors, see Eq.~\eqref{eq:FTphases} and \ref{app:A}, determines how finely we are able to resolve modulations of the FT phase factors (and hence the SC phase) in real space. If vortices are very close to materials lattice sites, these modulations in FT or SC phases become very strong.
Therefore, if one includes too few reciprocal lattice vectors in the FT phase calculation, an erroneous particle-hole symmetry breaking of the obtained spectrum occurs.
Performing tests similar to the ones in \ref{app:A}, we thus determine the maximum allowed deviations of vortex positions from plaquette centers for a given fixed and numerically feasible cutoff of reciprocal lattice summations.
\\

This leads us to restrict the disordered vortex lattice positions to the vicinity of plaquette centers as follows: for each vortex position $\Vr_j$ in Eq.~\eqref{eq:vortexpos}, we determine which plaquette of the materials lattice it is in. We may call this the ``plaquette location'' $\VbR_j'$ of the vortex. Note that the latter can be distinct from the original (non-disordered) vortex location $\VbR_j$ if the displacement $\bm{w}_j$ was sufficiently large. We then calculate the embedded vortex position
\begin{equation}\label{eq:vortexpos-embed}
\Vr_j' = \VbR_j' + \alpha (\Vr_j-\VbR_j')~,
\end{equation}
with $0\leq\alpha\leq 1$. For $\alpha=1$, the embedded vortex position is simply the bare disordered position $\Vr_j$, while for $\alpha=0$ we force the disordered lattice positions to also be centered at plaquettes, $\Vr_j' = \VbR_j'$. In the latter case, positional disorder hence only comes in multiples of the materials lattice constant. We find that numerically feasible reciprocal lattice summation cutoffs give us sufficient accuracy in the FT phase calculation to set $\alpha = 0.6$, while keeping erroneous PH symmetry breaking effects small enough to not adversely affect our results or conclusions.
All vortex positions and lattices in the following section hence will be embedded in the sense that vortex positions are at most $\alpha\cdot 0.5 = 0.3$ linear distance away from plaquette centers, i.e. each vortex keeps a linear distance of at least $0.2$ from materials lattice sites and edges.
In fact the vortex positions and reciprocal lattices in Fig.~\ref{fig:lattice_dis} were already generated by this prescription, and reflect the embedded vortex lattices used henceforth. Effects both due to the use of periodic boundary conditions (PBC) in a finite lattice and due to the vortex embedding are visible in Fig.~\ref{fig:lattice_dis}. Besides the overall hexagonal pattern in reciprocal space, one can see faint horizontal and vertical ripples which are due to the square lattice.\\

Finally we note that in Section~\ref{sec:MajTB} we will introduce a Majorana-only model that effectively simulates the lowest-energy modes of the vortex lattice. That model does not explicitly simulate the underlying material, and thus none of the above embedding issues apply, discussed here and for the non-disordered lattice case in Sec.~\ref{sec:setup-params}. For that reason the Majorana-only model is much easier to set up and analyze, both conceptually and numerically.

\subsection{Results for the disordered vortex lattice}
\label{sec:TISC_triag_dis}

\begin{figure*}[hbtp]
    \centering
    \includegraphics[width=\textwidth]{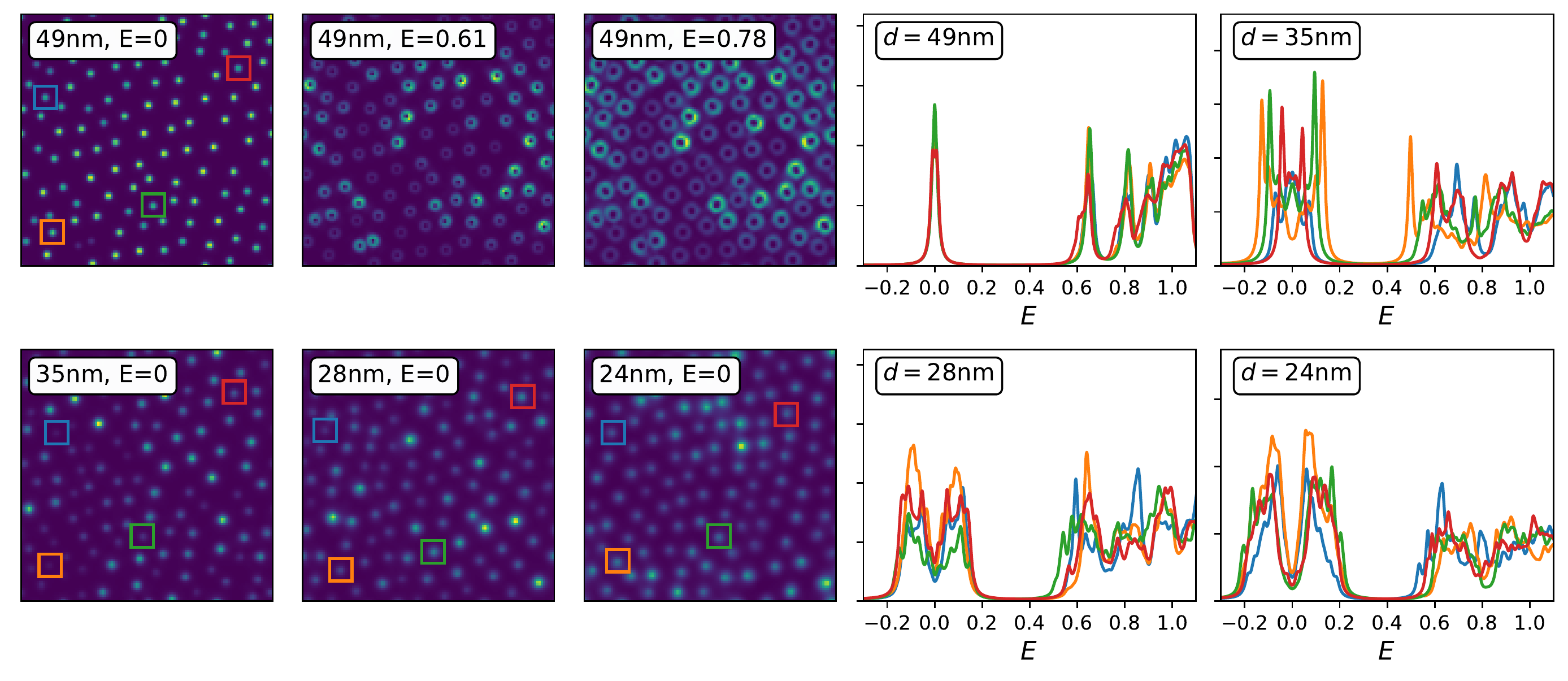}
    \caption{
    Spatial DOS and frequency-resolved LDOS for the $L=112$ and $n_v=112$ vortex lattice with varying average vortex spacing $d_v$, cf. Fig.~\ref{fig:LDOS_overview}, but now for the disordered lattice with $\sigma_d = 0.2$ and $\lambda_d = 2$, cf. Sec.~\ref{sec:lattice-dis} and Fig~\ref{fig:lattice_dis}.
    At each vortex separation, in the zero-energy DOS plot (left), we select four individual vortices for which we calculate the LDOS by integration over the respective colored box. The resulting LDOS plots (right) are color-coded accordingly.
    The top three plots (left) show the DOS for $d_v \approx 49 {\rm nm}$ and at energies of the first three peaks in the LDOS (right). Vortices again are well-separated and the zero-energy spectral peaks stay sharp, however some higher-energy modes are weakly split and show a variance in the LDOS of distinct vortices. In the DOS, some CdGM modes show a reduced intensity since their spectral weight is shifted and more spread out in energy.
    Upon decreasing the vortex distance $d_v$, we observe that peak splitting and disorder effects become significantly more pronounced, in particular for the lowest-energy modes.
    For $d_v \approx 35 {\rm nm}$, some vortices appear to be gapped while others have a zero-energy peak. In constrast, for $d_v \approx 28 {\rm nm}, 24 {\rm nm}$ all vortices are gapped, but the LDOS for these gapped low-energy modes and also that of the CdGM modes fluctuates strongly.
    Out of the vortex separations considered, it appears that $d_v \approx 35 {\rm nm}$ shows the strongest variation in zero-energy peak intensity in the DOS plots.
    }
    \label{fig:LDOS_overview_dis}
\end{figure*}

We here simulate near-triangular vortex lattices with positional disorder parametrized by the disorder variance $\sigma_d$ and correlation length $\lambda_d$ (quoted in units $d_v$), cf. Sec.~\ref{sec:lattice-dis}.
As for the regular triangular lattice in Sec.~\ref{sec:TISC_triag} and Fig.~\ref{fig:LDOS_overview}, we first investigate the DOS and LDOS for a few typical vortex separations $d_v \approx 49\,{\rm nm},~35\,{\rm nm},~28\,{\rm nm},~24\,{\rm nm}$. For illustration purposes, we here use strong disorder with $\sigma_d = 0.2$ and $\lambda_d = 2$, cf. Fig.~\ref{fig:lattice_dis} or the spatial DOS plots in Fig.~\ref{fig:LDOS_overview_dis} from which one can easily identify the vortex positions.
Also note that though PH symmetry on the level of individual vortices generally is broken, below we focus on the positive-frequency part of the LDOS and spatially resolved DOS. The behavior at negative frequencies is qualitatively similar.

The main observations of these simulations are discussed in Fig.~\ref{fig:LDOS_overview_dis}. Briefly, the disorder effects in the spatial DOS and vortex LDOS become more dramatic as the vortex separation $d_v$ is lowered. This is expected, since the hybridization energy scales increase and fluctuations in the latter can be resolved once they surpass the spectral broadening $\eta$. Also at low vortex separations the disorder conspires with the lattice frustration effects that lead to side-peaks for Majorana and CdGM modes, and produces a more complex multi-peak LDOS.
Finally, the low-energy peaks in LDOS and the zero-energy spatial DOS are most sensitive to disorder when the corresponding clean system is near a node of the Majorana hybridization oscillation. This is the case for $d_v \approx 35\,{\rm nm}$ in Fig.~\ref{fig:LDOS_overview_dis}, where some of the vortices show zero-bias peaks in the LDOS, while others do not.
We attribute this to the combination of strong fluctuations of coupling strengths under weak variation of the inter-vortex distance, together with the overall sizable mode hybridizations at small distances (overcoming the spectral broadening $\eta$). As discussed in detail in Sec.~\ref{sec:MajTB}, the fact that the sign of the effective Majorana hybridizations fluctuates between distinct modes then is the most crucial ingredient to obtain a strong LDOS variation~\cite{KrausStern2011}.
Qualitatively, the results in Fig.~\ref{fig:LDOS_overview_dis} look more similar to the experimental results of Machida et al.~\cite{Machida2019} as compared to the non-disordered case, Fig.~\ref{fig:LDOS_overview}. This motivates us to further investigate the effects of positional disorder.\\

\begin{figure}[hbpt]
	\centering
	\includegraphics[width=0.9\columnwidth]{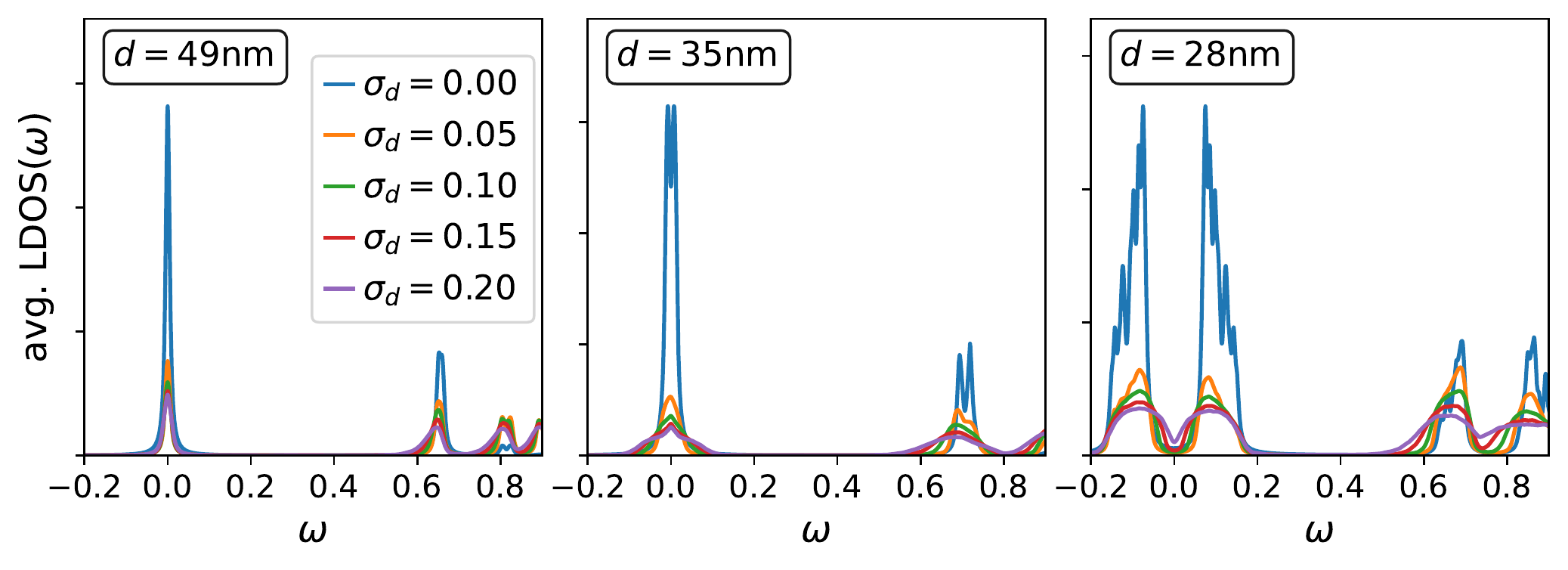}
	\caption{
	Average LDOS at vortex positions vs energy for the $n_v=112$ vortex lattice, at spacing $d_v\approx49{\rm nm},~35{\rm nm},~28{\rm nm}$ [$B = 1{\rm T},~2{\rm T},~3{\rm T}$]. Disorder parameters are $\lambda_d = 2$ and $\sigma_d \in [0, 0.2]$, and the broadening $\eta = 0.005\Delta_0$. Curves are obtained by averaging over $50$ disorder realizations and $n_v=112$ vortices per realization.
    If initially gapped, upon increasing $\sigma_d$ the system gradually becomes gapless and the LDOS develops a finite spectral weight at zero energy. How quickly this happens strongly depends on the value of the spacing~$d_v$.}
	\label{fig:TISCdos-avg}
\end{figure}

To make more qualified observations, we next consider disorder-averaged spectral functions. We here focus on the lowest-energy (most localized) modes and solve for ``only'' the $5\cdot n_v$ eigenstates closest to zero energy. Hence the overall spectral weight at high energies will be suppressed or altogether absent. In contrast, in Fig.~\ref{fig:LDOS_overview_dis} and to properly resolve higher-energy modes, we solved for significantly more eigenstates of the Hamiltonian which is correspondingly more expensive.
The results for three vortex separations $d_v\approx49{\rm nm},~35{\rm nm},~28{\rm nm}$ and various disorder strengths are shown in Fig.~\ref{fig:TISCdos-avg}. As for a single disorder realization in Fig.~\ref{fig:LDOS_overview_dis}, for a large vortex spacing $d_v\approx49{\rm nm}$ the disorder effect is hardly resolved, since the fluctuations of the vortex mode hybridizations are on the scale of the spectral broadening $\eta$.
When the average vortex spacing is close to a node in the MVM hybridization, $d_v\approx35{\rm nm}$, the initially split low-energy peaks merge and broaden as the disorder strength is increased, but a zero-energy ``hump'' remains. Hence, on average we would expect to find a peak at or close to zero energy more likely than not when scanning over an ensemble of vortices. Higher-energy CdGM modes that initially are split likewise merge into a smoothed out hump of spectral weight.
Finally, at a small average vortex spacing $d_v\approx28{\rm nm}$ where the low-energy modes are strongly split, disorder becomes ineffective in closing the gap. While some spectral weight is shifted towards zero energy, a distinct hump is left where the zero-disorder clusters of spectral peaks were located, cf. Fig.~\ref{fig:LDOS_overview}.
We again note that very similar behavior for the low-energy (Majorana) modes of the vortex lattice will be recovered in the simpler Majorana-only model, see Sec.~\ref{sec:MajTB} below.\\

\begin{figure}[hbpt]
	\centering
	\includegraphics[width=0.6\columnwidth]{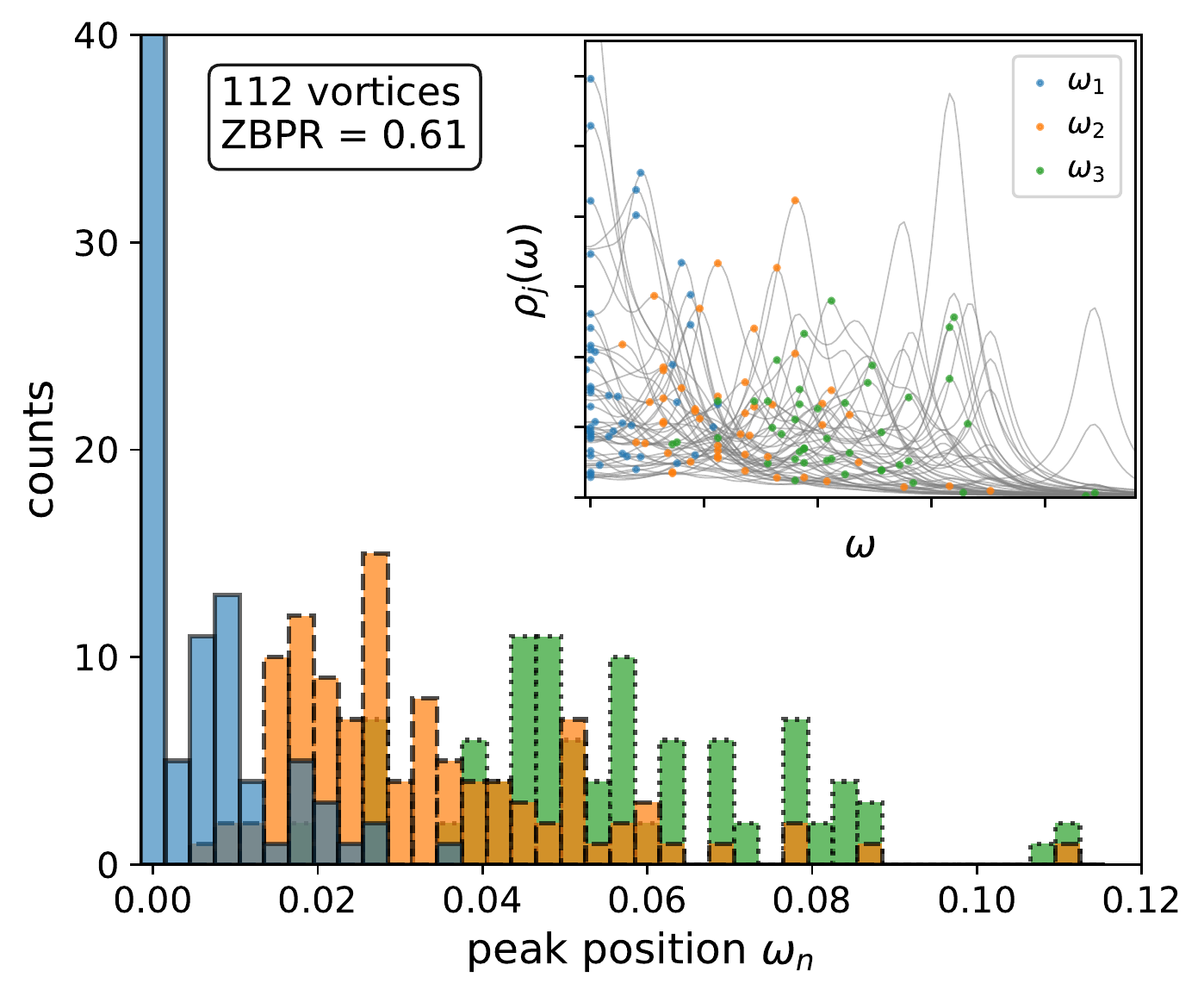}
	\caption{
	Analysis of LDOS spectra for the $n_v=112$ vortex lattice at spacing $d\approx35{\rm nm}$ ($B = 2{\rm T}$). Disorder parameters are $\sigma_d=0.2d$ and $\lambda_d=2d$ (cf. Fig.~\ref{fig:LDOS_overview_dis}), and STM resolution is $\eta = 0.005 \Delta_0$.
	Inset: spectra $\rho_j(\omega)$ for several modes, with first three peaks $\omega_{1,2,3}$ indicated by blue, orange, and green dots.
	Main: corresponding histogram of peak positions for LDOS spectra of all $n_v$ vortices, with a zero-bias peak rate of $\sim 0.61$.}
	\label{fig:TISCpeakstats}
\end{figure}

Motivated by the experimental analysis of Refs.~\cite{Machida2019,Chiu2020}, we finally consider the spectral peak statistics in the LDOS of vortex modes. I.e. we here inquire about an ensemble of vortices ($n_v=112$) that are in the ``field of view'' of our simulations (viz., the STM in experiment), and deduce the presence or absence of zero- and low-energy modes based on the locations of peaks in the vortex' LDOS. In Fig.~\ref{fig:TISCpeakstats} we consider a single vortex lattice and disorder realization, to illustrate how the peak statistics are obtained. First, we calculate the vortex LDOS for all $n_v$ vortices in the lattice, cf. Figs.~\ref{fig:LDOS_overview_dis} and \ref{fig:TISCdos-avg}.
The peaks in the vortex LDOS are identified simply by finding its local maxima, and peak heights and locations are indicated for a subset of vortex spectra in the inset of Fig.~\ref{fig:TISCpeakstats}. We then histogram the peak positions $\omega_n$, with the result shown in the main panel of Fig.~\ref{fig:TISCpeakstats}. Overall, after considering several disorder realizations, we find that the peak position distributions tend to be ``pyramid-shaped'', with the lowest spectral peaks clustering towards zero energy.
One quantity that often is quoted in STM studies of the vortex LDOS is the zero-bias peak rate (ZBPR), i.e., the fraction of vortices that shows a zero-energy peak. For the parameters considered here, we find ${\rm ZBPR} \approx 0.61$.
\begin{figure}[hbpt]
	\centering
	\includegraphics[width=0.8\columnwidth]{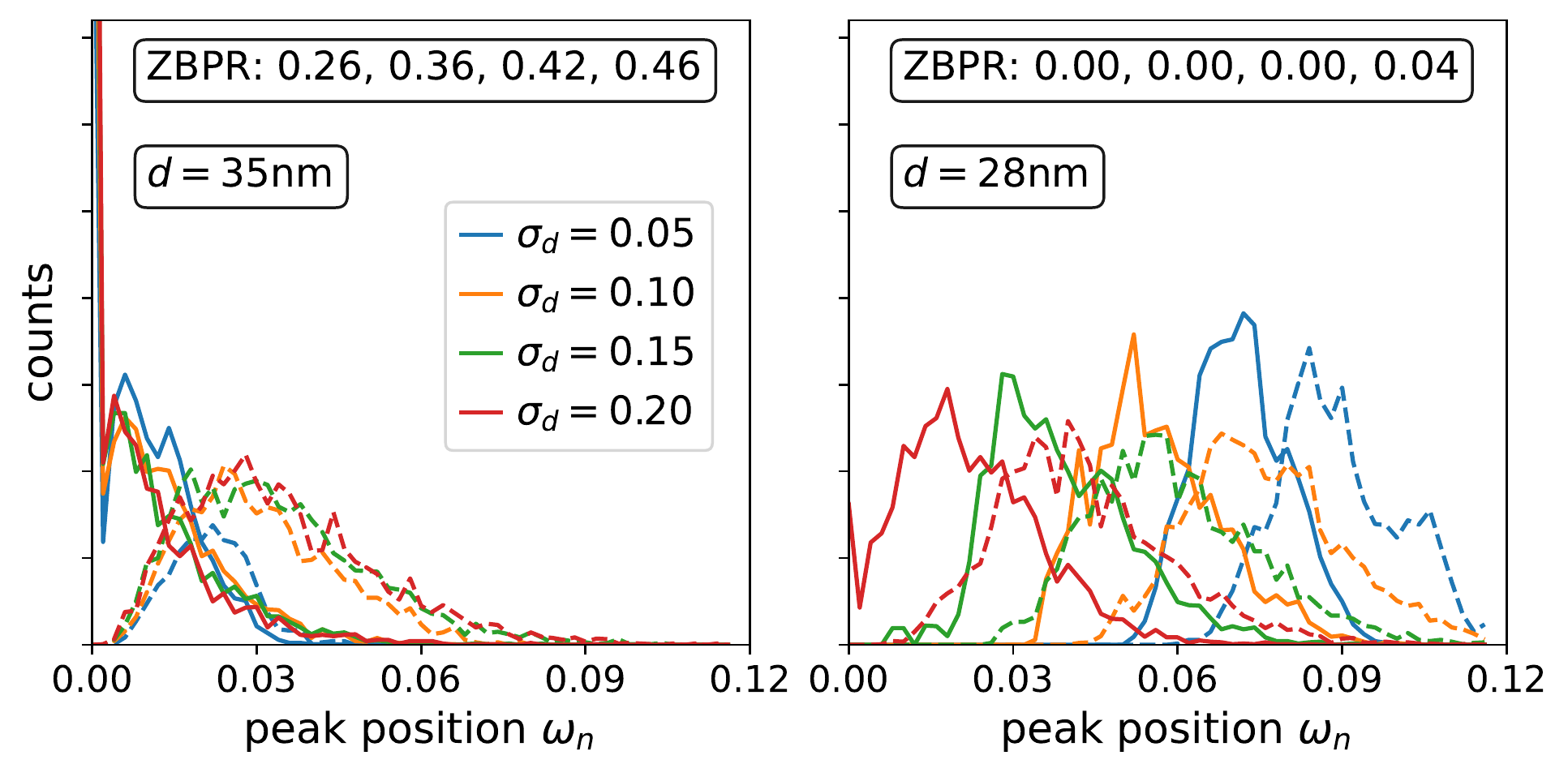}
	\caption{
	LDOS peak statistics for the $n_v=112$ lattice at spacing $d\approx 35{\rm nm},~28{\rm nm}$ ($B = 2{\rm T},~3{\rm T}$); parameters are $\lambda_d = 2$ and $\sigma_d \in [0.05, 0.2]$, and $\eta = 0.005 \Delta_0$. Data averaged over $50$ disorder realizations.
	Solid and dashed lines show peak positions $\omega_{1,2}$ respectively, cf. Fig.~\ref{fig:TISCpeakstats}.
	For an initial near-gapless system (left), increasing disorder $\sigma_d$ gradually increases the rate of low- and zero-energy peaks (ZBPR, box). In an initial gapped system (right), strong disorder is required to generate any zero-energy peaks.
    These observations are consistent with the data and discussion of Fig.~\ref{fig:TISCdos-avg}.}
	\label{fig:TISCpeakstats-avg}
\end{figure}

Beyond a single disorder realization, we consider the disorder-averaged LDOS peak statistics shown in Fig.~\ref{fig:TISCpeakstats-avg}. Here we find clear evidence for the pyramid-shaped distribution of LDOS peak positions.
As mentioned in the discussion of our non-disordered vortex lattice results, cf. Sec.~\ref{sec:TISC_triag} and Fig.~\ref{fig:LDOS_overview}, we can partially attribute them to the fanning out of the Majorana and CdGM bands in the triangular lattice model. Ultimately what is behind this is not the specific form of model or disorder, but rather the generic fact that (Majorana) fermions on a triangular lattice are frustrated.
As expected based on the behavior of the averaged LDOS, cf. Fig.~\ref{fig:TISCdos-avg}, the ZBPR increases significantly only when the non-disordered system already is close to a gapless state (for vortex spacing close to a node of the MVM hybridization oscilllation). If the system is initially strongly gapped, it takes rather strong disorder to generate any vortices that show zero-bias peaks in their LDOS. For $d_v\approx28{\rm nm}$, this means a disorder variance as large as $\sigma_d = 0.2$, see Fig.~\ref{fig:TISCpeakstats-avg}.
\\
We thus conclude this section by emphasizing that a key feature of vortex modes in experiments~\cite{Machida2019,Chiu2020}, i.e. pyramid-shaped low-energy peak distributions, are recovered in our model. In agreement with the analysis of Refs.~\cite{Machida2019,Chiu2020}, it was sufficient to include simple positional disorder of vortices as opposed to chemical or other disorder effects.

\section{Majorana-only  model}
\label{sec:MajTB}

A simplified approach to the disordered vortex lattice problem can be formulated in terms of a Majorana tight-binding model that considers only low-lying (Majorana) degrees of freedom. This is defined by the Hamiltonian
\begin{equation}\label{maj00}
H = i\sum_{j,k} t_{jk}\gamma_j\gamma_k ~,
\end{equation}
where $\gamma_j=\gamma_j^\dag$ denotes the Majorana operator representing the zero mode at vortex position ${\bf r}_j$ and $t_{jk}$ is a real antisymmetric matrix encoding the associated tunneling amplitudes.
In appropriate limits, i.e. at energies well below the CdGM modes ($\omega \ll\Delta^2/\mu$) and not too high density of vortices, this simpler model should qualitatively reproduce earlier results of Sec.~\ref{sec:FeSC} and Ref.~\cite{Chiu2020}.
As a major benefit, the implementation of the tight-binding problem is transparent and straight-forward, and allows us to easily consider thousands of Majorana modes (vortices) in the lattice.
For useful references on practical aspects in the implementation of Majorana lattice models, see Refs.~\cite{LiuFranz2015,KrausStern2011,Laumann2012,GrosfeldStern2006,Biswas2013,RahmaniFranz2019,Chiu2015} that are used extensively in this section.
Given the relative simplicity of a Majorana-only model, it also lends itself towards future research on interaction effects in disordered Majorana lattices~\cite{RahmaniFranz2019,Chiu2015,Affleck2017,Li2018,Rahmani2019,Tummuru2020}.

\subsection{Setup of the Majorana lattice model}
\label{sec:MajTB-setup}

Here we show how to set up the basic Majorana lattice model, to which one can introduce a tunable degree of positional disorder as discussed in Sec.~\ref{sec:lattice-dis}.
A simple way to implement a triangular lattice is to choose a two-site unit cell of ``$a$'' and ``$b$'' Majorana vortex modes (MVMs). Their relative positions in the unit cell are given as
\[
{\bm \delta}_a = (0,0)~~,~
{\bm \delta}_b = (dx/2,dy/2)~,
\]
with $dx = 1$ and $dy=\sqrt{3}$. The lattice translation vectors $\VbR_x = (dx,0), \VbR_y = (0,dy)$ span a rectangular lattice of this two-site unit cell. MVM positions are thus labelled by two integers, $n_x$ and $n_y$, and their ``flavor'' $a$ or $b$:
\begin{equation}\label{eq:unitcell}
\Vr = n_x\VbR_x+n_y\VbR_y+{\bm \delta}_{a/b}~.
\end{equation}
Aside from the form and choice of disorder, cf. Sec.~\ref{sec:lattice-dis}, there is only one ingredient to the model. The Majorana hybridization $t_{jk}$ between MVMs localized at $\Vr_j$ and $\Vr_k$ follows from the well-established analytical form~\cite{Cheng2009,Chiu2015,LiuFranz2015,Chiu2020}
\begin{equation}\label{eq:tjk-MVMs}
t_{jk} = t_{\rm hyb}(\Vr_{jk}) = t_{\rm mat}(r_{jk}) \sin(\omega_{jk})~,
\end{equation}
where $t_{\rm mat}(r_{jk})$ denotes a material-dependent Majorana hybridization, and the $\sin(\omega_{jk})$ factor captures geometric vortex lattice phases. Note that $t_{\rm mat}$ depends only on the relative distance $r_{jk} = |\Vr_j-\Vr_k|$ between the MVMs.
We first specify the materials-dependent part, cf. Eq.~\eqref{eq:MVMhyb},
\begin{equation}\label{eq:tjk-amp}
t_{\rm mat}(r_{jk}) = t_0\frac{\cos(k_F r_{jk}+\theta)}{\sqrt{r_{jk}}} e^{-r_{jk}/\xi}~,
\end{equation}
with Fermi momentum $k_F$, a phase $\theta$, and coherence length $\xi$. The bare tunneling amplitude $t_0$ sets the overall energy scale of the model; we henceforth put $t_0 = 2\Delta_0$ ($=3.8$\,meV), cf. the discussion in Sec.~\ref{sec:TISC_triag}. This will allow us to make quantitative comparisons with results in Sec.~\ref{sec:FeSC}. According to Table~\ref{table:params}, we further take $k_F^{-1} = 5{\rm nm}$, $\theta = \pi/4$ (or $\theta \approx 1.4$), and $\xi = 13.9{\rm nm}$.
Below we measure all distances in nanometer (${\rm nm}$), and scale the lattices with the characteristic inter-vortex distance $d$. Note that $d = 15-80{\rm nm}$ in experiments \cite{Chiu2020,Machida2019}.

\begin{figure}[hbpt]
	\centering
	\includegraphics[width=0.7\columnwidth]{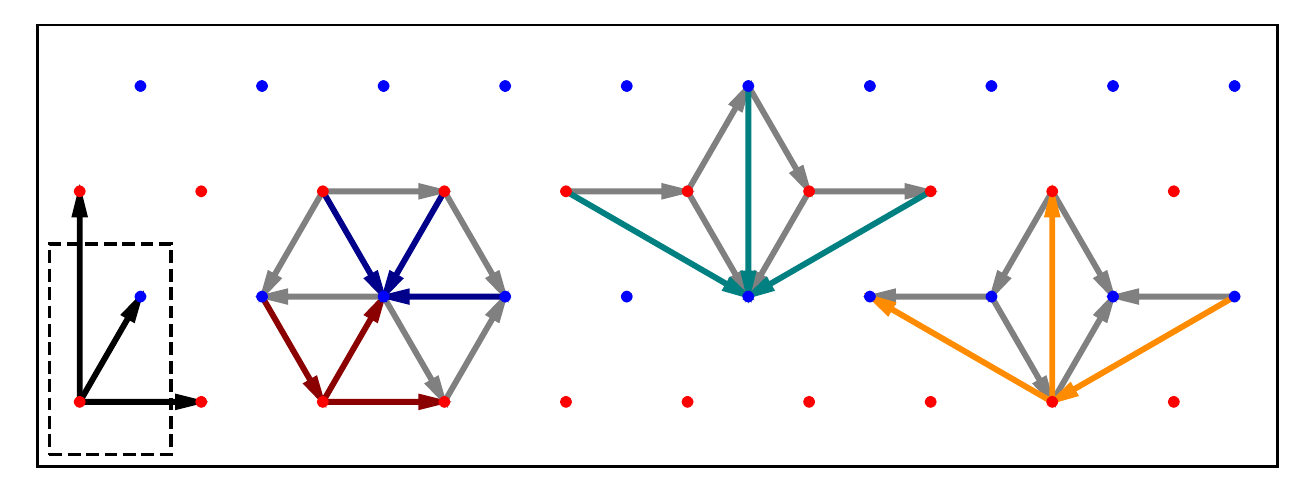}
	\caption{Triangular lattice, represented as a rectangular Bravais lattice with two-site unit cell (dashed box) of $a$ (red) and $b$ (blue) MVMs. Black arrows indicate the displacement vector between $a$ and $b$ MVM as well as lattice translation vectors. Colored arrows indicate GS phases $\pm i$ assigned when tunneling along links with/against the direction of the arrow~\cite{GrosfeldStern2006,LiuFranz2015}. Left: for tunneling between neighboring MVMs; red and blue arrows indicate all unique tunneling phases, while grey arrows are related by lattice translations. Going counter-clockwise around any plaquette of the lattice, one consistently obtains a phase of $+i$ indicative of a physical flux $\pi/2$ piercing that area. This gives the correct flux $\pi$ per MVM, each taking up two triangles of the lattice. Right: orange and teal arrows indicate next-neighbor hoppings with associated phases; these follow from neighbor hopping phases, by considering the grey arrows. Again each loop of two gray plus one colored arrows gives a total phase $+i$, consistent with the enclosed flux.
	}
	\label{fig:GSrule}
\end{figure}

Now consider the geometric vortex lattice phases, entering via $\omega_{jk}$ in Eq.~\eqref{eq:tjk-MVMs}.
The simplest way to evaluate $\omega_{jk}$ is by following the Grosfeld-Stern (GS) rule~\cite{GrosfeldStern2006}, which strictly only holds for regular lattices like the triangular or square lattice cases considered in Refs.~\cite{LiuFranz2015,Chiu2015}. Here we assume that the GS rule also applies in moderately disordered lattices, and discuss possible deviations. Taking the results of Refs.~\cite{LiuFranz2015,Chiu2015} and including next-neighbor hopping, we obtain the phase assignments shown in Fig.~\ref{fig:GSrule}.

For disordered lattices and without including phase factors $\omega_{jk}$ beyond the GS rule, we have to argue whether we can trust our numerical results at least qualitatively. Inspecting Eq.~\eqref{eq:tjk-MVMs}, the main effect of phases $\omega_{jk}$ different than those in the regular lattice will be to further modulate and randomize couplings $t_{jk}$. Also local fluctuations of physical parameters $k_F$, $\theta$, and $\xi$ will tend to randomize the couplings $t_{jk}$. Introducing just one source of randomness and one way in which it enters, namely the positions of MVMs and their separations entering $t_{jk}$, might underestimate the effect of disorder. Conversely, to capture the experimentally observed effects, we might have to increase the disorder strength to larger values than naively expected.
Still this allows for a significant computational simplification since calculating the tunneling amplitudes only requires knowing the distances between MVMs, Eq.~\eqref{eq:tjk-amp}, while to calculate $\omega_{jk}$ exactly one has to consider the full MVM lattice arrangement.
We have also repeated the analysis of the disordered triangular lattice as in Sec.~\ref{sec:lattice-dis} and Fig.~\ref{fig:lattice_dis}, but now using the base triangular lattice defined by Eq.~\eqref{eq:unitcell}. While we here do not provide an additional figure, note that since for the Majorana-only model we are able to go to significantly larger vortex lattices, the effects of employing PBC on a torus are much less visible. Further the embedding problem is completely avoided in the Majorana-only model, since there is no underlying materials lattice.\\

Let us also note that in earlier work, Kraus and Stern~\cite{KrausStern2011} and Laumann et al.~\cite{Laumann2012} considered the case of disordered tunneling amplitudes and phases in Majorana lattices; these were simply picked as random variables. Naturally it would be easy for us to do the same, and pick the phase values $\omega_{jk}$ according to some random distribution. However we do not expect to gain much additional insight from this.
Rather, here we take into account fluctuating MVM separations as motivated from and observed in the recent experiments~\cite{Machida2019,Chiu2020}. Still it will be instructive to compare our results with those obtained in Refs.~\cite{KrausStern2011,Laumann2012}.

\subsection{MVM lattice simulation and observables}
\label{sec:MajTB-sim-obs}

With the setup of the Majorana lattice in Secs.~\ref{sec:MajTB-setup} and \ref{sec:lattice-dis}, Figs.~\ref{fig:GSrule} and \ref{fig:lattice_dis}, we now calculate GS phase factors and tunneling amplitudes according to Eq.~\eqref{eq:tjk-MVMs}. It is useful to rewrite the Majorana tight-binding Hamiltonian \eqref{maj00} as
\begin{equation}
H = i\hat{\gamma}^T \hat{h} \hat{\gamma}~,
\end{equation}
with Majorana vectors $\hat{\gamma}^T = (\gamma_1,...,\gamma_{2N})$, and $\hat{h}$ a real anti-symmetric matrix with entries $t_{jk}$ in Eq.~\eqref{eq:tjk-MVMs}.
After a Schur decomposition of the matrix $\hat{h}$ into real, anti-symmetric $2\times2$ blocks on its diagonal~\cite{Kitaev2001}, one obtains
\begin{equation}\label{eq:HmvmDiag}
H =  i\tilde{\gamma}^T \tilde{h} \tilde{\gamma} = \sum_{m=0}^{N-1} 2\varepsilon_m i\tilde{\gamma}_{2m}\tilde{\gamma}_{2m+1}~,
\end{equation}
where $\tilde{h} = O\hat{h}O^T$ and $\tilde{\gamma} = O\hat{\gamma}$ with orthogonal matrix $O$.
The pairs $\tilde{\gamma}_{2m},~\tilde{\gamma}_{2m+1}$ then define $N$ complex fermions $c_m = \tilde{\gamma}_{2m} + i \tilde{\gamma}_{2m+1}$, and $H = \sum_m \varepsilon_m (c_m^\dagger c_m-\frac12)$.
One can now bin and plot the energy eigenvalues $\varepsilon_m$, averaged over disorder realizations or with varying $\sigma_d$ and $\lambda_d$, to compare with the earlier results of Refs.~\cite{KrausStern2011,Laumann2012}.
\\

For us more interestingly, we can obtain the Greens and spectral functions for the original Majoranas $\gamma_j$ by using the above transformation $O$. The local spectral density of $\gamma_j$ is what is measured by an STM tunneling experiment~\cite{Machida2019,Kong2019,Kong2020} into the MVM and vortex at position $\Vr_j$.
Let us consider the Lehmann representation of the retarded Majorana Greens function (GF), at zero temperature,
\begin{equation}\label{eq:GFlehmann}
G_j^R(\omega) = \sum_n\left[\frac{|\bra{n}\gamma_j\ket{0}|^2}{\omega+E_0-E_n+i\eta} + (E_0\leftrightarrow E_n)\right]~,
\end{equation}
where $\ket{n}$ is the many-body (MB) eigenstate at energy $E_n$, and $\ket{0}$ is the ground state.
To relate to eigenvalues $\varepsilon_m$, we write $\gamma_j$ via $\tilde{\gamma_j}$ in terms of complex fermions $c_m$,
\[
\gamma_j = \sum_{m=0}^{N-1} (\psi_{mj}c_m + \psi_{mj}^\ast c_m^\dagger)~.
\]
Here $\psi_{mj} = \frac12(O_{2m,j}-iO_{2m+1,j})$ is derived from the above orthogonal transformation $O$.
Since we are dealing with a single-particle problem, the MB states are product states $\ket{n} = \ket{n_0\cdots n_{N-1}}$. For the MB ground state $\ket{0}$, each fermion is filled or empty depending on the sign of the corresponding energy $\varepsilon_m$. Applying $\gamma_j$ to $\ket{0}$, either $c_m$ or $c_m^\dagger$ thus annihilate the state. For the $m$th term in the sum, matrix elements in Eq.~\eqref{eq:GFlehmann} hence evaluate as
\[
|\bra{n}(\psi_{mj}c_m + \psi_{mj}^\ast c_m^\dagger)\ket{0}|^2 = |\psi_{mj}|^2 |\bra{n}c_m\ket{0} + \bra{n}c_m^\dagger\ket{0}|^2 .
\]
If we remove or add an electron from/to the ground state (first/second term), the energy is lowered/raised by $\varepsilon_m$, giving $E_0 - E_n = \mp \varepsilon_m$. Since we consider a particle-hole symmetric form for the GF, adding $E_0\leftrightarrow E_n$ in Eq.~\eqref{eq:GFlehmann}, energy denominators with either sign are present in the final result.
Hence the retarded Majorana GF for modes $\gamma_j$ reads
\begin{equation}\label{eq:GFmaj}
G_j^R(\omega) = \sum_m |\psi_{mj}|^2 \left[\frac{1}{\omega-\varepsilon_m+i\eta} + \frac{1}{\omega+\varepsilon_m+i\eta}\right].
\end{equation}
We thus obtain a symmetrized sum of retarded GFs of the systems' eigenmodes $c_m$, weighted by their amplitudes $|\psi_{mj}|^2$ at the MVM lattice site $\Vr_j$.
The associated local spectral function then follows as $\rho_j(\omega) = -\frac1\pi {\rm Im} G_j^R(\omega)$.
Note that in constrast to the LDOS and spatial DOS in the TI-SC model, cf. Sec.~\ref{sec:LDOSdef}, the Majorana LDOS as defined here is necessarily PH symmetric.
\\

We can now calculate the local spectra $\rho_j(\omega)$ for large lattice sizes ($N = L_xL_y$ unit cells, $2N$ MVMs, PBC), varying the mean vortex distance $d$ (magnetic field $B$) and disorder parameters $\sigma_d$ and $\lambda_d$. The broadening $\eta$ here mimics the finite spectral resolution of the STM measurement in experiments, as in Sec.~\ref{sec:FeSC}.
Depending on parameters, we then check how many of the MVMs have zero-bias peaks (zero-bias peak rate, ZBPR), what is the distribution of the lowest and higher-energy peak positions across all vortices, and so on. Having access to large lattices with thousands of MVMs allows us to obtain better statistics than is possible with the more detailed models of Sec.~\ref{sec:FeSC} and Ref.~\cite{Chiu2020}.
This approach thus should enable us to reproduce some of the experimental observations~\cite{Machida2019,Kong2019,Kong2020,Chen2018}, and provide valuable information on whether the low-energy physics in iron-based SCs can be understood in terms of Majorana vortex modes.

\subsection{Results for the Majorana lattice model}

\begin{figure}[hbpt]
	\centering
	\includegraphics[width=0.9\columnwidth]{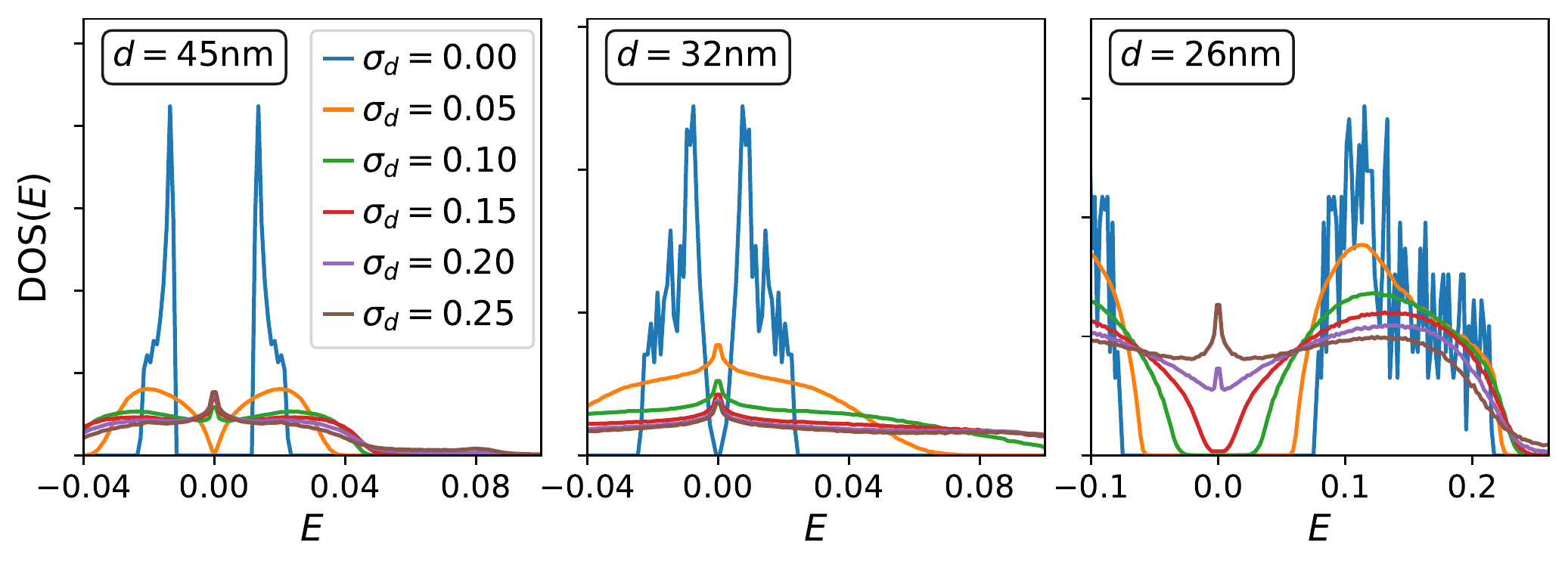}
	\caption{
	DOS vs energy for a MVM lattice with $2N = 1840$ modes at spacing $d=45{\rm nm},~32{\rm nm},~26{\rm nm}$ ($B \simeq 1{\rm T},~2{\rm T},~3{\rm T}$); disorder parameters are $\lambda_d = 2$ and $\sigma_d \in [0, 0.25]$. Each curve is averaged over $500$ disorder realizations.
    With increasing $\sigma_d$, the system gradually becomes gapless and the DOS develops a peak at zero energy. How quickly this happens strongly depends on the value of the spacing $d$, for discussion see text.
    }
	\label{fig:MVMdos}
\end{figure}

Here we discuss some results from the MVM lattice simulations outlined in Secs.~\ref{sec:MajTB}A-B.
The single-particle density of states (DOS) of the model is shown in Fig.~\ref{fig:MVMdos}. We obtain it by binning the energies $\varepsilon_m$ in Eq.~\eqref{eq:HmvmDiag}, with an average over disorder realizations for fixed parameters $\sigma_d$ and $\lambda_d$, cf. Sec.~\ref{sec:lattice-dis}, and fixed spacing $d$.

Comparing to results of Kraus and Stern~\cite{KrausStern2011}, we find that the DOS in Fig.~\ref{fig:MVMdos} are qualitatively similar to their random-hopping models with nearest-neighbor terms. While we have a more regular structure of couplings, derived from the underlying near-triangular lattice, this difference does not seem to affect the density of states too much.
However we do expect the vortex spacing $d$ to play an important role, especially for weak disorder when the cosine-modulation of couplings in Eq.~\eqref{eq:tjk-amp} is prominently visible.
To this end, note that parameters $d$ and $\sigma_d$ translate into typical vortex separations $r \in (d-\sigma_d,d+\sigma_d)$. Entering these into Eqs.~\eqref{eq:tjk-MVMs}-\eqref{eq:tjk-amp}, we notice that if the average vortex spacing $d$ is tuned close to a node of the oscillating MVM hybridization, couplings $t_{jk}$ will be random both in sign and amplitude. Instead if $d$ is close to a local maximum (or minimum), introducing a small variance $\sigma_d$ only scrambles the amplitude but not the sign of $t_{jk}$'s.
As shown in Ref.~\cite{KrausStern2011}, the two cases are quite different: random-sign models are gapless and show a peak of the DOS at zero energy, while with random amplitude (but fixed sign) the model stays gapped. We hence expect a gapped system, without zero-bias peaks for the local DOS at vortex cores (see below), whenever $d$ is far from a node of the MVM hybridization and disorder is weak (small $\sigma_d$ and/or large $\lambda_d$). When $d$ is close to a node, even weak disorder will randomize the signs of couplings and lead to a gapless system with zero-modes in many vortices. Finally, for strong disorder (large $\sigma_d$), the precise value of the spacing $d$ does not matter anymore and typical vortex separations $r\in(d-\sigma_d,d+\sigma_d)$ lead to random-sign couplings and hence a gapless system.
Indeed the results shown in Fig.~\ref{fig:MVMdos} reflect these trends. Comparing to results in the TI-SC model of Sec.~\ref{sec:TISC_triag_dis}, cf. Fig.~\ref{fig:TISCdos-avg}, we find that its low-energy modes show very similar behavior as the Majorana-only model in Fig.~\ref{fig:MVMdos}.\\

\begin{figure}[hbpt]
	\centering
	\includegraphics[width=0.6\columnwidth]{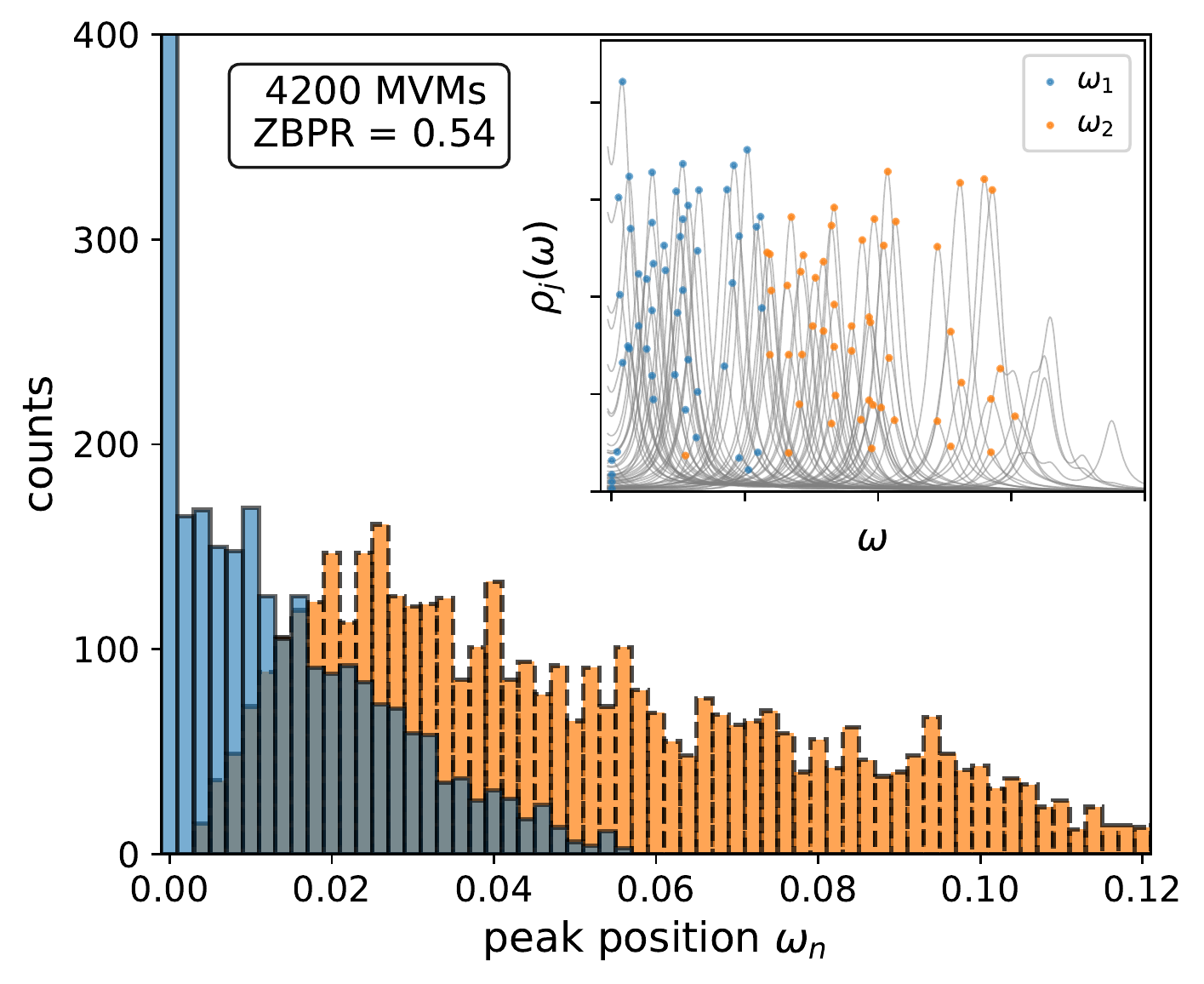}
	\caption{
	MVM spectra for a vortex lattice with $2N = 4200$ modes at spacing $d=32{\rm nm}$ ($B \simeq 2{\rm T}$). Disorder parameters are $\sigma_d=0.2d$ and $\lambda_d=2d$ (cf. Fig.~\ref{fig:lattice_dis}), and STM resolution is $\eta = 0.002 t_0$.
	Inset: spectra $\rho_j(\omega)$ for several modes, with first and second peaks $\omega_{1,2}$ indicated by blue and orange dots.
	Main panel: corresponding histogram of peak positions for all $2N$ modes, with a zero-bias peak rate of $\sim 0.54$. Note the pyramid shape of the peak distributions, cf. also Refs.~\cite{Machida2019,Chiu2020}.
	}
	\label{fig:MVMpeakstats}
\end{figure}

Next we consider the LDOS and spectral peak statistics of MVMs, cf. Sec.~\ref{sec:TISC_triag_dis} and Refs.~\cite{Machida2019,Chiu2020}. As discussed in Sec.~\ref{sec:MajTB-sim-obs} and shown in Fig.~\ref{fig:MVMpeakstats}, calculating the LDOS allows to extract spectral peak frequency histograms and the zero-bias peak rate (ZBPR). When the system becomes gapless and consecutively develops a zero-bias peak in the total DOS of Fig.~\ref{fig:MVMdos}, it also shows a non-zero ZBPR for the LDOS of MVMs. Moreover, we find that the lowest and second-lowest energy peak positions $\omega_{1,2}$ generically are distributed in a ``pyramid-shape'' way in Fig.~\ref{fig:MVMpeakstats}, in accordance with observations in experiment~\cite{Machida2019} and the more complex models of Sec.~\ref{sec:FeSC} and Ref.~\cite{Chiu2020}.\\

\begin{figure}[hbpt]
	\centering
	\includegraphics[width=0.8\columnwidth]{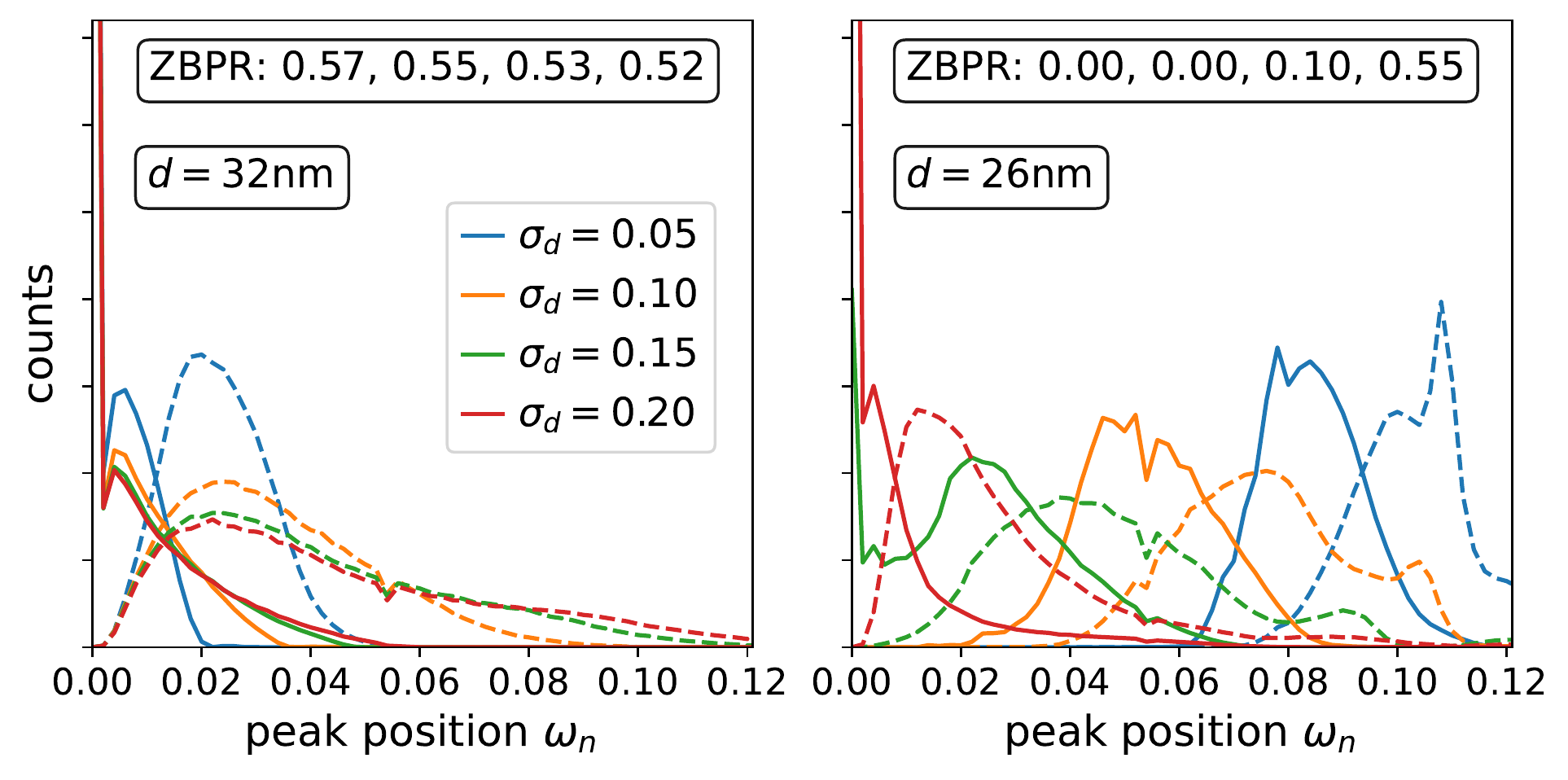}
	\caption{
	MVM peak statistics for a vortex lattice with $2N = 1840$ modes at spacing $d= 32{\rm nm},~26{\rm nm}$ ($B \simeq 2{\rm T},~3{\rm T}$); parameters are $\lambda_d = 2$ and $\sigma_d \in [0.05, 0.2]$, and $\eta = 0.002 t_0$. Data is averaged over $500$ disorder realizations.
	Solid and dashed lines show peak positions $\omega_{1,2}$ respectively, cf. Fig.~\ref{fig:MVMpeakstats}.
	For an initial near-gapless system (left), after jumping to $\sim 0.5$ the ZBPR (text box) slowly decreases under stronger disorder. In an initial gapped system (right), increasing $\sigma_d$ gradually increases the rate of low- and zero-energy peaks.
	In both cases, increasing disorder causes the peak positions to spread out across a larger range of energies.
	These observations are consistent with the data and discussion of Fig.~\ref{fig:MVMdos}.}
	\label{fig:MVMstats-avg}
\end{figure}

Beyond a single disorder realization in Fig.~\ref{fig:MVMpeakstats}, we again take disorder averages to analyze trends in the ZBPR and peak statistics upon varying the disorder parameters $\sigma_d$, $\lambda_d$ and spacing $d$. This data is shown in Fig.~\ref{fig:MVMstats-avg}.\\

The MVM peak statistics observed in Fig.~\ref{fig:MVMstats-avg} closely agree with our intuition gained from Fig.~\ref{fig:MVMdos}, and also with the more complex model in Sec.~\ref{sec:TISC_triag_dis} and Fig.~\ref{fig:TISCpeakstats-avg}. Generally, disorder broadens spectral features and smears out the distributions of spectral peak positions in Fig.~\ref{fig:MVMstats-avg}.
The overall trends depend on whether the system initially, in the absence of disorder, is strongly gapped or close to a gapless state (spacing $d$ far from or close to a node of the oscillating MVM hybridization, Eq.~\eqref{eq:tjk-amp}).
If the initial system is near-gapless, disorder quickly closes the gap and leads to an accumulation of spectral peaks at zero energy. Ramping up $\sigma_d$ then spreads out the spectral weight and actually decreases the ZBPR, which is different from the full TI-SC model in Fig.~\ref{fig:TISCpeakstats-avg}. This may be due to the absence of higher-energy CdGM modes in the Majorana-only model, which otherwise would provide a source from which spectral weight also is transferred to lower energies.
For a strongly gapped system, the threshold to generate a finite ZBPR is larger, but the ZBPR increases even for strong disorder. This is more similar to the behavior in the full TI-SC model.
Finally at very strong disorder, say $\sigma_d>0.2d$, the cosine-modulation of the MVM hybridization Eq.~\eqref{eq:tjk-amp} becomes inessential. The main trend that determines the ZBPR then is the decay $\langle t_{jk} \rangle_{\sigma_d,\lambda_d} \sim e^{-d/\xi}/\sqrt{d}$ of the hybridizations $t_{jk}$ with increasing vortex spacing $d$.\\

\begin{figure}[hbpt]
	\centering
	\includegraphics[width=0.6\columnwidth]{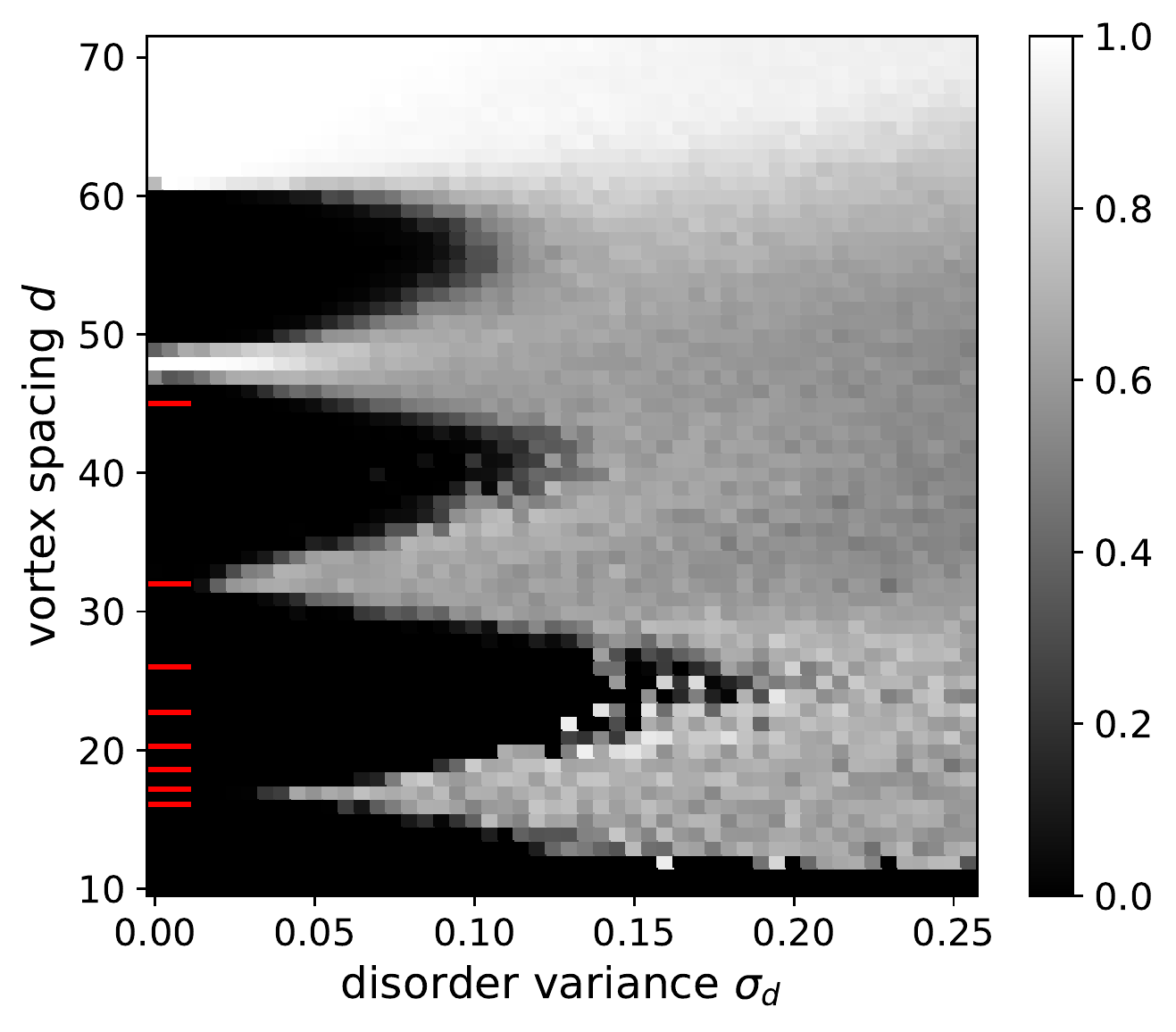}
	\caption{
	Zero-bias peak rate versus disorder strength $\sigma_d$ and spacing $d$ for a vortex lattice with $2N = 1840$ modes. Data is averaged over $50$ disorder realizations. The disorder correlation length is $\lambda_d=2d$, and the broadening $\eta = 0.002 t_0$. The red bars indicate spacings $d= 45{\rm nm},32{\rm nm},...,16.1{\rm nm}$, corresponding to magnetic field strengths $B \simeq 1{\rm T},2{\rm T},...,8{\rm T}$. At weak to moderate disorder, the ZBPR clearly follows the underlying MVM hybridization oscillations. At strong disorder, only the general trend to lower ZBPR at smaller $d$ is retained.
    }
	\label{fig:MVM-ZBPR}
\end{figure}

Last, we analyze how the ZBPR varies as function of the spacing $d$ and disorder strength $\sigma_d$. The extraction of the ZBPR here proceeds analogous to Figs.~\ref{fig:MVMpeakstats} and~\ref{fig:MVMstats-avg}, and results are shown in Fig.~\ref{fig:MVM-ZBPR}. Naturally the ZBPR also depends on the disorder correlation length $\lambda_d$ and the peak broadening $\eta$ (``STM resolution''), and we here pick values that are convenient for illustration purposes.

The ZBPR in Fig.~\ref{fig:MVM-ZBPR} for small to intermediate $\sigma_d$ shows clear signatures of the hybridization oscillations, Eq.~\eqref{eq:tjk-amp}. Whenever the MVM hybridization is strong the ZBPR is suppressed, and vice versa. For stronger disorder $\sigma_d \gtrsim 0.15d$ and large spacing $d\simeq 30-60{\rm nm}$ that suppresses the overall energy scales in the system, the ZBPR quickly averages to around $0.4-0.6$. The trend of increasing ZBPR with increasing spacing $d$, at strong disorder, is clearly observed in Fig.~\ref{fig:MVM-ZBPR}.\\

Finally, let us relate back to experimental findings. Machida et al.~\cite{Machida2019} observe an increase in vortex lattice disorder as the magnetic field is ramped up. In terms of our model, we hence should consider an increasing disorder variance $\sigma_d$ with fixed or decreasing disorder correlation length $\lambda_d$ as the vortex spacing $d$ is reduced. In Fig.~\ref{fig:MVM-ZBPR}, this means taking some line cut that starts at large $d$ and small $\sigma_d$, and ends at small $d$ and large $\sigma_d$.
Likewise one could investigate the effect of a changing broadening $\eta$, reflecting the STM resolution in experiment~\cite{Machida2019,Wang2018,Chen2018,Kong2019,Kong2020,Zhang2018}. In this section we set $\eta=0.002t_0$, which according to our discussion in Figs.~\ref{fig:MVMhyb_2v} and \ref{fig:MVMhyb_triag} translates to $\eta = 0.004\Delta_0 \approx 7.2\,\mu eV$. This is close to the experimentally quoted $\eta_{\rm STM} \approx 20\mu eV$. Also note that Ref.~\cite{Machida2019} employs a more sophisticated peak-fitting procedure than simply checking for maxima in the local spectral density, which effectively enhances their spectral peak resolution to a larger value. Last and more importantly, our simulations are implemented in the zero-temperature limit; here one could also include finite temperature, in form of an additional Gaussian broadening of the spectral peaks.

\section{Conclusion and Outlook}
\label{sec:conclusion}

Motivated by several recent experimental studies of low-energy vortex bound states in the iron-based superconductor FeTe$_{0.55}$Se$_{0.45}$~\cite{Zhang2018,Wang2018,Machida2019,Kong2019,Kong2020,Chen2018}, in this work we investigated Majorana and Caroli-de Gennes-Matricon (CdGM) vortex modes that arise in periodic vortex lattices hosted on the surface of proximitized topological insulators.
To this end, we discussed a Fu-Kane type square lattice model~\cite{Fu2008,Marchand2012,Pikulin2017} that captures the low energy physics of the surface state of a topological insulator proximity-coupled to an $s$-wave superconductor, and obviates the need to simulate bulk degrees of freedom~\cite{Chiu2020}.
A modified version of the \FranzTesanovic singular gauge transformation~\cite{Franz2000,Vafek2001} then allows us to impose arbitrary vortex arrays in the TI-SC model under periodic boundary conditions which avoids complications arising from gapless modes that would be associated with the samples' edges.
\\

After checking that our model correctly reproduces the analytical prediction for the Majorana vortex mode hybridization~\cite{Cheng2009}, we applied it to the case of regular and disordered triangular vortex lattices. As seen in the more complex model of Ref.~\cite{Chiu2020}, we find that both the spatially resolved density of states at fixed energies and local spectra of individual vortices indeed exhibit several features that were observed in experiment~\cite{Machida2019,Chiu2020}.
At large vortex lattice spacings (small magnetic fields), a large fraction of vortices exhibits low- or zero-energy spectral peaks. By increasing the applied magnetic field, thus decreasing the average vortex spacing, the hybridization between nearby vortices is increased. Even in the non-disordered case, the Majorana and CdGM modes that are hosted in vortex cores then exhibit a ``fanning out'' of the associated spectral weight in the vortex LDOS. By comparison with a Majorana-only model, we confirm that this is not an artifact but a physical effect occuring for Majorana fermions on a frustrated lattice.
In combination with configuration disorder of the vortex array, at intermediate or small average vortex separations, this leads to ``pyramid-shaped'' distributions of peak positions in the vortex LDOS. By sampling the vortices in a given lattice configuration and averaged over disorder realizations, we thus find an accumulation of spectral weight (and peaks) towards zero energy. Such statistics of spectral peak positions were analyzed in experiment~\cite{Machida2019} and recovered also in the more complex model of Chiu et al.~\cite{Chiu2020}.
Where overlapping, our results broadly agree with the experimental results of Ref.~\cite{Machida2019} and theoretical analysis of Ref.~\cite{Chiu2020}. However our models are significantly simpler than that of Chiu et al.~\cite{Chiu2020}. Our description thus potentially allows to go to much larger vortex lattice sizes, and to include effects of interactions, impurities, or quasi-particles. Likewise we could imagine simulating the complex vortex lattice geometries that might be of interest towards Majorana manipulation and topological quantum information processing applications.
\\

There are many interesting experimental setups and future research avenues that can be tackled by means of the modified \FranzTesanovic transformation. Recall that this method relies only on the effective BdG Hamiltonian description of the superconducting lattice model under consideration.
A present direction of interest is the simulation of patterned superconducting island arrays that proximitize the two-dimensional electron gas hosted in an InAs quantum well. High-quality samples of such systems recently were grown to serve as a potential platform for 2D Majorana device arrays~\cite{Prada2020review,Lutchyn2018review,Boettcher2018}. In this context, B{\o}ttcher et al.~\cite{Boettcher2018} observed a superconductor-metal-insulator transition that is tunable by electrostatic gates and subject to flux commensurability effects in a perpendicular applied magnetic field.
While some aspects of the devices' response were understood in terms of effective scaling theories that apply to SC island arrays~\cite{Kapitulnik2019RMP}, a better microscopic description of its physics clearly is desirable if such setups proceed to be used towards the realization of topological quantum computation.
We note that Ref.~\cite{Levine2017} predicted that such Al-InAs superconducting island arrays in an in-plane magnetic field may realize 1D or 2D topological superconducting phases. However, the latter work neglected orbital (out-of-plane) effects of the magnetic field. Also the results of B{\o}ttcher et al.~\cite{Boettcher2018} left open several questions as to what is the effect of in-plane fields on transport in 2D arrays, where they observed changed scaling behavior that might be indicative of a transition away from a purely 2D phase.
\\

Further, as an extension of the Majorana lattice model, one can mimic the presence of finite-energy CdGM modes by adding auxiliary coupled Majorana modes at each lattice size. To this end, define two additional MFs $\chi_j$ and $\tilde{\chi}_j$ for each original Majorana $\gamma_j$. The auxiliary modes are ``dangling'' from the lattice in the sense that they only locally couple to $\gamma_j$ (and to each other), with a hybridization $\kappa \sim \Delta_0^2/\mu$ reflecting the expected CdGM mode energy.
In the local spectral functions of the original modes $\gamma_j$ this leads to additional satellite peaks, and hence to side-bands in the lattice model, at the correct CdGM mode energies. Coupling additional auxiliary Majoranas at each site may also give a PH-asymmetric LDOS.
Note that this way of adding the lowest-energy CdGM modes still results in a simple, sparse Majorana tight-binding model, where the number of involved Majorana fermions and hence the Hamiltonian matrix size has ``only'' tripled. Hence one still should be able to easily obtain corresponding results as shown in Sec.~\ref{sec:MajTB} even for the extended model.
\\
As a technically more challenging extension, we again mention the interacting disordered Majorana lattice models~\cite{RahmaniFranz2019,Chiu2015,Affleck2017,Li2018,Rahmani2019,Tummuru2020}. These might be of interest, e.g., towards the quantification of decoherence and decay channels for quantum information stored in Majorana vortex arrays.
\\

Finally, while here we focused on a positional disorder of the vortex lattice, it will also be worthwhile to consider the effect of spatially varying material parameters. Somewhat surprisingly, the analysis of Machida et al.~\cite{Machida2019} indicated that vortex locations and the absence or presence of zero-modes do not seem to correlate with variations of the superconducting gap or the local chemical composition of the underlying material. It would be useful to understand why that is the case.
To this end, if one imposes a spatially varying SC pairing strength, one should probably solve the vortex lattice problem self-consistently. In a system with homogeneous SC pairing, as considered in the present work, it is known that vortices arrange in a triangular lattice in order to minimize the SC free energy. This will be different if the SC pairing strength fluctuates spatially, since vortices may find local weak-pairing regions in which they can be accommodated more cheaply. 
Conversely, one might argue that a spatially varying SC pairing strength could be what gives rise to the disordered vortex lattice in the first place (though Ref.~\cite{Machida2019} seems to indicate otherwise).

\subsection*{Acknowledgements}
We acknowledge useful discussion with P. Lopes, E. Lantagne-Hurtubise, O. Can, C.K. Chiu, and T. Liu. This work was supported by
NSERC, the Max Planck-UBC-UTokyo Centre for Quantum Materials and the Canada First Research Excellence Fund, Quantum Materials and Future Technologies Program.

\appendix

\section{Modified FT transformation}\label{app:A}

Here we provide technical details regarding the modified \FranzTesanovic (FT) singular gauge transformation~\cite{Franz2000,Vafek2001} for vortex lattices in SC systems under PBC.

\subsection{Calculation of phase factors}\label{a1}

First, Maxwell's equation for the magnetic field in a superconductor with super-current ${\bm j}$ is
\begin{equation}\label{maxwell}
\begin{split}
    \nabla\times\bB&=\mu_{0}{\bm j}\\
    &=\mu_{0}n_{s}e{\bm v}_{s},
\end{split}
\end{equation}
where $\mu_{0}$ is the permeability, and $n_{s}$ the carrier density.
The super-fluid velocity ${\bm v}_{s}$ is defined as
\begin{equation}
    {\bm v}_{s}=\frac{\hslash}{m}\left(\frac{1}{2}\nabla\phi(\Vr)-\frac{e}{\hslash c}\bA\right).
\end{equation}
Taking the curl of Eq.~\eqref{maxwell} and using $\nabla\cdot\bB=0$, we get the London equation in the presence of vortices,
\begin{equation}\label{london}
    \bB-\lambda_{s}^{2}\nabla^{2}\bB=\frac{\Phi_{0}}{2}\hat{z}\sum_{j}\delta(\Vr-\Vr^{j}),
\end{equation}
where the $j$ summation runs over all vortices (both of type A and B), $\lambda_{s}=\sqrt{\frac{mc}{\mu_{0}n_{s}e^{2}}}$ is the London penetration depth and $\Phi_{0}=\frac{hc}{e}$ is the SC flux quantum.
\\

We define superfluid velocities for the A and B vortex groups as $\bv_{g}=\frac{\hslash}{m}\left(\nabla\phi_{g}-\frac{e}{\hslash c}\bA\right)$ with $g=A,B$ such that $\bv_{s}=\frac{\bv_{A}+\bv_{B}}{2}$. The phase factors in Eq.~\eqref{phases} are expressed in terms of superfluid velocities as $\mathcal{V}_{g}^{\alpha}(\Vr)=\frac{m}{\hslash}\int^{\Vr+\alpha}_{\Vr}\bv_{g}\cdot\bm{dr}$. Using Eq.~\eqref{london} and following the steps as in Refs. \cite{Franz2000,Vafek2001}, we get the expression for the superfluid velocity $\bv_{g}$ as
\begin{equation}\label{vg}
    \bv_{g}=\frac{2\pi\hslash}{m}\int\frac{d^{2}k}{(2\pi)^{2}}\left(i\bm{k}\times\hat{z}\right)\left[\frac{\sum_{j}\textrm{e}^{i\bm{k}.(\Vr-\Vr_{g}^{j})}}{\lambda_{s}^{-2}+k^{2}}\right].
\end{equation}
In this equation, it is implicit that the sum runs over all vortex positions $\Vr^{j}_{g}$ of type $g$ for an infinite lattice. The infinite lattice comprises magnetic unit cells of dimensions $L\times L$ repeating periodically in space. 
Taking $\bG=\frac{2\pi}{L}(n_{x},n_{y})$ as the reciprocal lattice vector of the magnetic unit, we can discretise the integral in the expression for the superfluid velocity, resulting in
\begin{equation} 
    \bv_{g}=\frac{2\pi\hslash}{m L^{2}}\sum_{\bG}\left(i\bG\times\hat{z}\right)\left[\frac{\sum_{j}\textrm{e}^{i\bG.(\Vr-\Vr_{g}^{j})}}{\lambda_{s}^{-2}+G^{2}}\right]
\end{equation}
The FT phase factors are then calculated as, 
\begin{equation}\label{FTphases}
     \mathcal{V}^{\alpha}_{g}(\Vr)=\frac{2\pi}{ L^{2}}\sum_{\bG,n}\left[\frac{i\bG\times\hat{z}}{\lambda_{s}^{-2}+G^{2}}\cdot\int_{\Vr}^{\Vr+\alpha}\textrm{e}^{i\bG\cdot(\Vr-\Vr^{n}_{g}})\bm{dr}\right].  
\end{equation}
In principle, the sum in Eq.~\eqref{FTphases} runs over an infinite number of reciprocal lattice vectors $\bG$. In practice, we employ a soft cutoff of the type $e^{-\alpha\vert G\vert^{2}/G_{\textrm{max}}^{2}}$, where $G_{\textrm{max}}$ is the maximum norm of reciprocal vectors and $\alpha>0$ is a constant. The soft cutoff reflects the vortex core size, and is necessary because the London model used to calculate the superfluid velocity assumes a constant SC order parameter amplitude. This is no longer true close to the vortex center.

\subsection{Apparent symmetry breaking under PBC}
\label{sec:symmetry}

The application of PBC results in subtleties related to the assignment of vortices into A and B sublattice. Here we review two symmetry constraints that must be respected by any physical model, but are broken if one naively applies the FT recipe~\cite{Franz2000,Marinelli2000,Vafek2001} as introduced above.

\subsubsection{Internal gauge symmetry}

In the formulation of the FT transformation, the choice of A and B vortex sublattice is entirely arbitrary. For a magnetic unit cell with two or more vortices, upon exchanging the A and B assignment of vortices, physical observables hence must be invariant. Irrespective of the choice of A-B configuration, for example, the spectrum should remain the same. As one can convince oneself, this works fine in the case of two vortices per unit cell. To check the internal gauge symmetry explicitly for a larger number of vortices, we choose a system with four vortices per magnetic unit cell. There are two distinct choices of A-B sub-lattices, dubbed `ABAB' and `AABB', as illustrated in Fig.~\ref{fig:ABsub-lattice}b-c.
The internal gauge symmetry now should be preserved in the lattice version of the BdG Hamiltonian~\cite{Vafek2001}, cf. Eq.~\eqref{eq:pos_space_bdg}.
To test this, we calculate the energy eigenvalues for the ABAB and AABB configurations after constructing the Hamiltonian with the phase factors in Eq.~\eqref{eq:FTphases}. As shown in Fig.~\ref{fig:4v_original}, the two configurations do not exhibit the same eigenvalues. Rather the difference between the two configurations is significant ($\approx10^{-2}$), much larger than any numerical errors. This suggests that for a finite lattice system and under PBC, the internal gauge symmetry is violated if the FT transformation is performed as introduced above~\cite{Franz2000,Vafek2001,Marinelli2000}.
We will have to reconcile and resolve this issue below.

\subsubsection{Particle-hole symmetry}

\begin{figure}[hbtp]
    \centering
    \includegraphics[width=0.6\columnwidth]{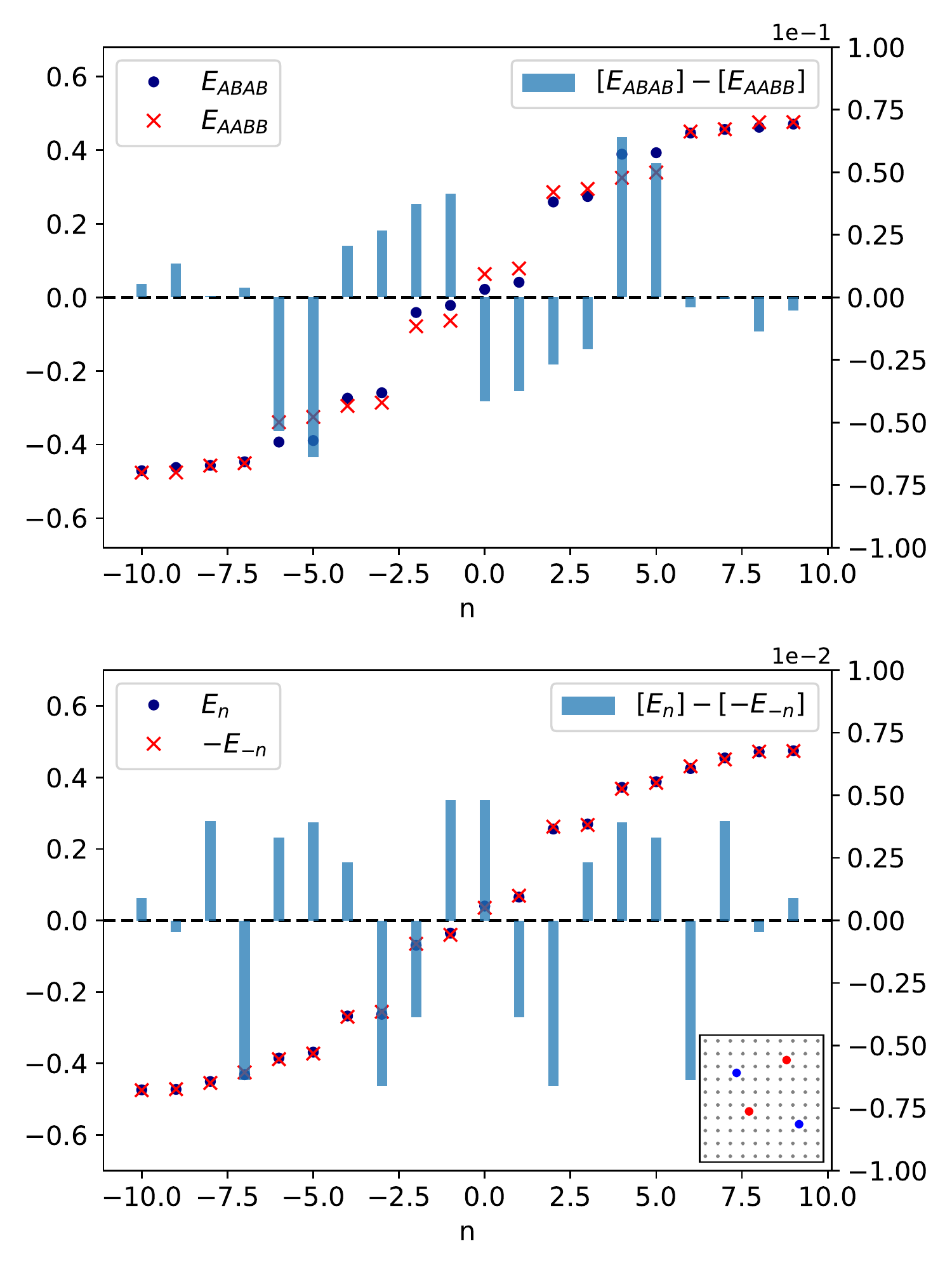}
    \caption{
    Top: energy eigenvalues for a lattice with two equivalent labels of its four vortices, ABAB and AABB in Fig.~\ref{fig:ABsub-lattice}b-c, used in the naive FT transformation of Sec.~\ref{sec:FT}. Evidently, the internal gauge symmetry relating the two ways of labelling is broken. The difference between the energy eigenvalues is of the order $\sim 10^{-1}$ and is depicted in the superimposed bar graph. 
    Bottom: energy eigenvalues for a vortex lattice with four vortices (inset; A-type in red, B-type in blue).
    The spectrum clearly does not exhibit PH symmetry, as the eigenvalues (blue) and their TR counterpart (red) do not match. The PH breaking error is of order $\sim 10^{-2}$, see the bar graph. Parameters for the simulations here are $\Delta_{0}=0.2$, $\mu=-0.45$, $\lambda=1$, and $m=0.5$. However, for the observed effects, the exact parameter choice is unimportant.
    }
    \label{fig:4v_original}
\end{figure}

By construction, the BdG Hamiltonian is particle-hole symmetric.
However in the FT transformation Eq.~\eqref{eq:U}, we have assigned distinct phases $\phi_A$ and $\phi_B$ to the particle and hole blocks of the BdG Hamiltonian.
As we will see now, for multiple vortices per unit cell this makes the PH symmetry considerations more complicated. For the four-vortex configurations in Fig.~\ref{fig:ABsub-lattice}b-c the spectrum is PH symmetric, irrespective of whether the spectra for ABAB and AABB labelling match. If we introduce disorder by moving vortices away from their original high-symmetry positions in the magnetic unit cell, see inset of Fig.~\ref{fig:4v_original}, the PH symmetry breaks down. The in-gap energy eigenvalues for such a disordered configuration are shown in the main panel of Fig.~\ref{fig:4v_original}. They clearly exhibit strong deviations from PH symmetry.

The spectrum in the BdG formalism is PH symmetric if the Hamiltonian of holes is the time-reversed counterpart of the Hamiltonian of particles. For a random vortex arrangement, if the Hamiltonian of particles and Hamiltonian of holes are not related by an anti-unitary operator, the PH symmetry breaks. Hence, the PH symmetry breaking is related to the choice of gauge in the theory developed until now.
Again, we will have to revise the FT transformation, and resolve this issue of apparent PH symmetry breaking.

\subsection{Modified FT transformation}\label{sec:revised_ft}

\begin{figure}[hbtp]
    \centering
    \includegraphics[width=0.6\columnwidth]{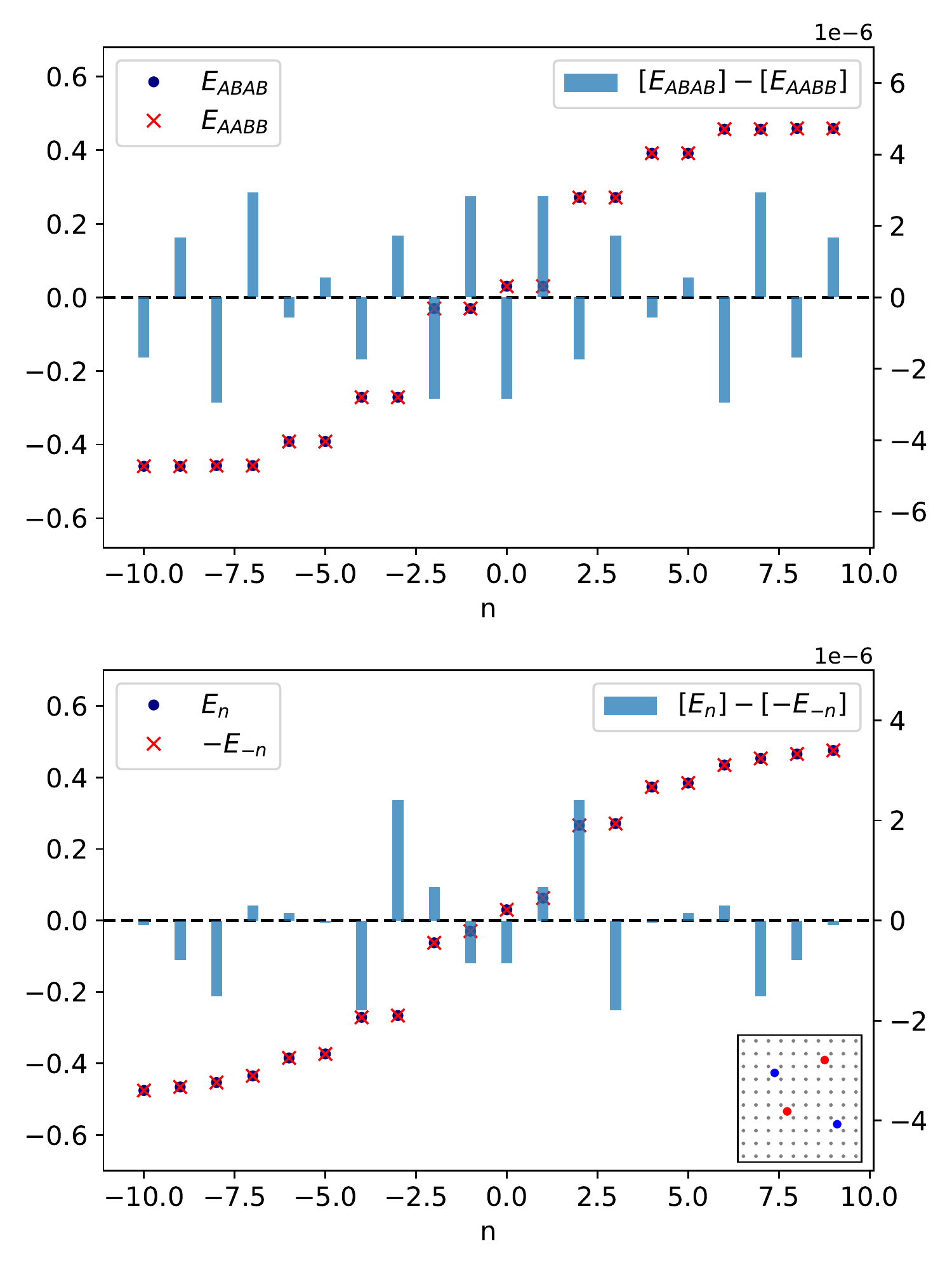}
    \caption{Resolution of internal gauge symmetry (A/B label) and PH symmetry issues. Top: using the modified FT phases now gives the same spectrum irrespective of the choice of A/B label (up to an acceptable error of the order $\sim 10^{-6}$, due to the reciprocal lattice sum cutoff). Bottom: Energy eigenvalues for a distorted vortex configuration (inset), again using the modified FT phases. The eigenvalues (blue) and their TR counterpart (red) now match within a small error ($\sim 10^{-6}$). Parameters are the same as in Figs.~\ref{fig:4v_original}}.
    \label{fig:4v_modified}
\end{figure}

The key to reconcile the apparent PH asymmetry and the dependence of observables on the choice of A/B label lies in understanding the physical constraint governing the choice of the A/B sub-lattice. Electrons and holes on the underlying physical lattice must experience the same magnetic field. For this constraint to hold,  the flux threaded through any plaquette of the lattice that is felt by electrons and holes can only differ by multiples of $2\pi$. This property  indeed is built to the FT transformation when it is applied to an infinite system.
Upon imposing PBC on the FT-transformed BdG Hamiltonian, we now explicitly need to ensure that the fluxes through the two additional, non-contractible loops in the torus geometry (2D lattice with PBC) obey the same principle.
This is in order to obtain a physical lattice with PBC, on which particles and holes experience the same magnetic fields even when traversing around the full circumference of the torus.
\\

Let us begin with the Bloch-transformed BdG Hamiltonian at wave vector $\bm{k}$, i.e.
$\mathcal{H_{\bm{k}}} = e^{-i\bm{k}\cdot\bm{r}}U^{-1} \mathcal{H}U e^{i\bm{k}\cdot\bm{r}}$.
Tunneling terms in the particle (A) and hole (B) blocks then acquire modified FT phase factors, which read
\begin{equation}\label{modification}
\mathcal{\Tilde{V}}_{A/B}^{\alpha}(\Vr) = 
\mathcal{V}_{A/B}^{\alpha}(\Vr) \pm k_{\alpha}\cdot\alpha,
\end{equation}
where $\alpha=\delta x,~\delta y$ are the hopping directions on the lattice. The problem of defining a PH-symmetric model now translates to finding a wave vector $\bm{k}$ for which electrons and holes are subject to the same flux through the non-contractible loops $C_{1,2}$ on the torus.
We then use this wave vector to amend the PBC of the lattice model for both sectors, thus re-casting it to correctly simulate the $\Gamma$-point $\bm{k} = (0,0)$ or any other wave vector we wish to investigate.
Recalling that the FT phase factor is related to the superfluid velocity as $\mathcal{V}_{g}^{\alpha}(\Vr)=\frac{m}{\hslash}\int_{\Vr}^{\Vr+\alpha}\bv_{g}(\Vr)\cdot\bm{dr}$ with $g=A,B$ (cf.~\ref{a1}), we obtain the condition
\begin{equation}
\oint_{C} \left(\frac{m}{\hbar}\bv_{A}(\Vr)+\bm{k}\right)
\cdot\bm{dr} = \oint_{C} \left(\frac{m}{\hbar}\bv_{B}(\Vr)-\bm{k}\right) \cdot\bm{dr}~.
\end{equation}
With $\Vr_{0}=(0,0)$ as the origin, the integral over loop $C_{1}$ runs from $\Vr_{0}=(0,0)$ to $\Vr_{1}=(L,0)$, and the integral over loop $C_{2}$ runs from $\Vr_{0}=(0,0)$ to $\Vr_{2}=(0,L)$. We here take $L_x = L_y = L$.
After integration and further simplification, we obtain the desired wave-vector $\bm{k}$ as
\begin{equation}\label{kxky_final}
\bm{k}=\frac{\pi}{L}\cdot\frac{\Delta\Vr\times\hat{\bm{z}}}{L} ~.
\end{equation}
Here $\Delta \Vr=\sum_{j\in A}\Vr_A^j - \sum_{j\in B}\Vr_B^j$, with $j$ running over the vortices of type A and B, respectively.
The flux that must be threaded through the two non-contractible loops in order to retain a PH-symmetric model can now be evaluated as $\phi_{\alpha}=\int_{0}^{L}k_\alpha d\alpha$, with $\alpha=x,y$. This gives $\phi_\alpha=\pi\frac{\Delta r_\alpha}{L}$. Implementing these changes then gives us modified FT phases valid for SC systems under PBC that correctly produce PH symmetric spectra.\\

Let us now revisit the issues identified in Figs.~\ref{fig:4v_original}-\ref{fig:4v_modified} and \ref{sec:symmetry}. After incorporating the correction terms in the FT phases, we correctly find spectra that respect the internal gauge symmetry (A/B label) and PH symmetry, see Fig.~\ref{fig:4v_modified}. 
The reason why the internal gauge symmetry issue is also resolved by the above discussion is that a relabeling of vortices, cf. Fig.~\ref{fig:ABsub-lattice}, shifts the positions of A- and B-type vortices. Hence, relabelling can also affect the flux threaded through the loops on the torus for electrons and holes, if the latter was not correctly compensated in the first place; i.e., the internal gauge symmetry breaking was just PH symmetry breaking in disguise.
We can thus use the expressions quoted in Sec.~\ref{sec:FT}, see also Eqs.~\eqref{modification} and~\eqref{kxky_final}, as a modified FT transformation that correctly describes SC systems under PBC.

\bibliography{tisc}
\end{document}